\documentclass[preprint,12pt]{elsarticle}

\biboptions{sort&compress}

\usepackage{graphicx}%
\usepackage{multirow}%
\usepackage{amsmath,amssymb,amsfonts}%
\usepackage{amsthm}%
\usepackage{mathrsfs}%
\usepackage[title]{appendix}%
\usepackage{xcolor}%
\usepackage{textcomp}%
\usepackage{manyfoot}%
\usepackage{booktabs}%
\usepackage{algorithm}%
\usepackage{algorithmicx}%
\usepackage{algpseudocode}%
\usepackage{listings}%
\DeclareMathOperator*{\argmaxA}{arg\,max}
\theoremstyle{thmstyleone}%
\theoremstyle{thmstyletwo}%

\theoremstyle{thmstylethree}%
\begin{document}

\begin{frontmatter}



\title{Enhanced PINNs for data-driven solitons and parameter discovery for (2+ 1)-dimensional coupled nonlinear Schrödinger systems}


\author[1]{Hamid Momeni} 
\ead{momeni.umz@gmail.com}
\author[1]{AllahBakhsh Yazdani Cherati \corref{cor1}}
\ead{yazdani@umz.ac.ir}
\cortext[cor1]{Corresponding author}
\affiliation[1]{organization={Department of Applied Mathematics, Faculty of Mathematical Sciences},
            addressline={University of Mazandaran}, 
            city={Babolsar},
            country={Iran}}
 \author[2]{Ali Valinejad} 
\ead{valinejad@umz.ac.ir}       
 \affiliation[2]{organization={Department of Computer Sciences, Faculty of Mathematical Sciences},
            addressline={University of Mazandaran}, 
            city={Babolsar},
            country={Iran}} 

\begin{abstract}
This paper investigates data-driven solutions and parameter discovery to (2+ 1)-dimensional coupled nonlinear Schrödinger equations with variable coefficients (VC-CNLSEs), which describe transverse effects in optical fiber systems under perturbed dispersion and nonlinearity. By setting different forms of perturbation coefficients, we aim to recover the dark and anti-dark one- and two-soliton structures by employing an enhanced physics-based deep neural network algorithm, namely a physics-informed neural network (PINN). The enhanced PINN algorithm leverages the locally adaptive activation function mechanism to improve convergence speed and accuracy. In the lack of data acquisition, the PINN algorithms will enhance the capability of the neural networks by incorporating physical information into the training phase. We demonstrate that applying PINN algorithms to (2+ 1)-dimensional VC-CNLSEs requires distinct distributions of physical information. To address this, we propose a region-specific weighted loss function with the help of residual-based adaptive refinement strategy. In the meantime, we perform data-driven parameter discovery for the model equation, classified into two categories: constant coefficient discovery and variable coefficient discovery. For the former, we aim to predict the cross-phase modulation constant coefficient under varying noise intensities using enhanced PINN with a single neural network. For the latter, we employ a dual-network strategy to predict the dynamic behavior of the dispersion and nonlinearity perturbation functions. Our study demonstrates that the proposed framework holds significant potential for studying high-dimensional and complex solitonic dynamics in optical fiber systems.
\end{abstract}




\end{frontmatter}



\section{Introduction}\label{sec1}

In optical fiber communication, nonlinear differential equations are widely used to model light propagation behavior, accounting for dispersion, nonlinearity, and attenuation to accurately predict signal transmission. These equations help researchers understand complex light-fiber interactions, enabling the design of efficient and reliable communication systems. \cite{bib1,bib2,bib3,bib4,bib5,bib6,bib7,bib8,bib9,bib10,bib11,bib12,bib13,bib14}. In this context, the nonlinear Schrödinger equation (NLSE) and its extensions play a crucial role in understanding pulse dynamics and soliton formation in optical fibers. These equations have been widely studied in many scientific disciplines, including nonlinear optics \cite{bib15,bib16,bib17}, fluid dynamics \cite{bib18}, quantum mechanics \cite{bib19, bib20}, and plasma physics \cite{bib21,bib22}. 

To model more realistic systems, various extensions of the standard NLSE have been introduced, including the coupled NLSE (CNLSE), which describes pulse propagation in multi-mode or birefringent fibers. Furthermore, higher-dimensional versions, such as the (2+ 1)-dimensional CNLSE, have been employed to study complex spatio-temporal structures like vortex solitons and two-dimensional localized waves. However, real-world optical fibers often exhibit spatial or temporal variations due to inhomogeneities, imperfections, or external perturbations, which cannot be captured by constant-coefficient models. To address this, the variable-coefficient (2+ 1)-dimensional coupled NLSE (VC-CNLSE) has been proposed, in which the dispersion and nonlinearity coefficients vary with respect to time or space \cite{bib53,bib54,bib55,bib56}. The (2+ 1)-dimensional VC-CNLSE can be formulated as:
\begin{align} \label{eq:3}
i\psi_{1,t} &+ \alpha(t)(\psi_{1,xx} + \psi_{1,yy}) + \beta(t)(\lvert \psi_{1} \rvert^2 + \tau \lvert \psi_{2} \rvert^2)\psi_{1} = 0,\notag\\
i\psi_{2,t} &+ \alpha(t)(\psi_{2,xx} + \psi_{2,yy}) + \beta(t)(\tau \lvert \psi_{1} \rvert^2 + \lvert \psi_{2} \rvert^2)\psi_{2} = 0,
\end{align}
where $i$ denotes the imaginary unit, $\psi_{1}(x,y,t)$ and $\psi_{2}(x,y,t)$ are both complex functions, representing the amplitudes of wave propagation through a two-mode optical fiber, and the subscripts $x$, $y$, and $t$ denote the partial derivatives with respect to these variables. The parameter $\tau$ varies, $\tau >2/3$ and $\tau <7$, regarding the Kerr-shape electron nonlinearity and the nonlinearity resulting from molecular orientation changes. By choosing different forms of perturbation functions $\alpha(t)$ and $\beta(t)$, one can obtain various optical soliton solutions by controlling factors such as intensity, velocity, and acceleration. Many studies have investigated solutions to the VC-CNLSE \eqref{eq:3} in recent years. By virtue of the Hirota method, linear, parabolic, and quasi-periodic dark solitons were reported in \cite{bib57}. Abundant dark and anti-dark soliton structures, as well as their collisions, are studied in \cite{bib58} using the Hirota method. By utilizing the developed Hirota bilinear method, one-soliton, two-soliton, three-soliton, and breather-like bright-dark solitons are observed in \cite{bib59}.

With the growing importance of data analysis, machine learning has emerged as a powerful tool for extracting insights and making predictions from large datasets, with applications in natural language processing \cite{bib60}, image recognition \cite{bib61}, bioinformatics \cite{bib62,bib63}, and climate modeling \cite{bib64}. Deep learning, a machine learning paradigm that leverages the universal approximation capability of neural networks, automatically learns complex features from data through its multi-layered architecture. Its success lies in processing vast amounts of data to uncover intricate patterns that are not apparent to traditional methods \cite{bib65,bib66}. However, model performance depends heavily on data availability; insufficient data can lead to poor robustness and lack of convergence. Incorporating prior knowledge—such as domain constraints, expert input, or physics-based models—can guide solutions toward admissible spaces, especially where data acquisition is costly. Recently, deep learning has been successfully applied to solving differential equations, giving rise to the emerging field of scientific machine learning.

Physics-informed neural networks (PINNs) combine neural networks with physical laws to predict and simulate complex phenomena accurately \cite{bib67}. Unlike traditional numerical methods, PINNs embed governing differential equations into the training process as a form of regularization, thereby enhancing their approximation capabilities. They treat spatial and temporal coordinates as inputs and employ automatic differentiation (AD) \cite{bib68} for efficient derivative computation, enforcing physical constraints through a multi-objective loss that integrates both data-driven and physics-informed residuals. This framework ensures that predictions remain consistent with underlying physical principles. PINNs have found applications in fluid dynamics, materials science, and medical imaging, offering flexibility for nonlinear, irregular, or inverse problems where traditional methods often struggle. Nevertheless, challenges and limitations remain \cite{bib70,bib71}. Moreover, PINNs are relatively straightforward to implement, with similar coding structures for both forward and inverse problems.

In recent years, numerous studies have been reported on the applications of PINNs in the field of optical fiber communication. In \cite{bib72}, an adaptive residual points PINN scheme is introduced for vector soliton simulation and parameter discovery of general CNLSE. Data-driven vector soliton solutions and parameter discovery for the coupled mixed derivative NLSE are studied in \cite{bib73}, where a PINN method with a twin subnet strategy is used. In \cite{bib74}, an improved PINN algorithm is introduced to simulate both vector degenerate and nondegenerate solitons of the coupled nonlocal NLSE and to estimate the associated equation parameters under varying noise intensities. A PINN scheme with a locally adaptive activation function is employed in \cite{bib75} for data-driven localized wave solutions of the derivative NLSE. In \cite{bib76}, the standard PINNs is used for prediction of soliton dynamics and estimation of model parameters for (2+ 1)-dimensional NLSE. In \cite{bib766}, the authors employed an improved PINN algorithm for solving forward and inverse problems, where the model equation was the Hirota equation with variable coefficients. They applied the proposed algorithm for data-driven discovery of the constant coefficients and the linear function variable coefficients in the model equation. In \cite{Song2023}, the authors extend PINNs to model stationary and non-stationary solitons of $1D$ and $2D$ saturable nonlinear Schrödinger equations (SNLSEs) with PT-symmetric potentials. They also propose a modified PINNs (mPINNs) to directly identify potential functions from solution data, investigate inverse problems, and compare network structures. A residual-unit-based gradient-enhanced PINN (R-gPINN) for solving forward and inverse problems of variable-coefficient PDEs is proposed in \cite{Zhou2025}. The authors introduced residual units to address gradient vanishing and network degradation, and incorporates gradient terms of variable coefficients to better enforce physical constraints. Through numerical experiments on several equations (Burgers, KdV, Sine-Gordon, and KP), they show that R-gPINN significantly improves the accuracy of both solution prediction and coefficient discovery.

Most studies above focus on simulating waveform propagation, modeled by the NLSE and its extensions, in homogeneous environments. Traditional numerical methods struggle with high-dimensional problems due to the curse of dimensionality, leading to exponential increases in computational cost and complexity. For strongly nonlinear problems, they often face convergence issues and numerical instability, especially when coupled equations involve multi-scale interactions or solutions with sharp gradients and localized structures like solitons. These limitations have motivated the development of alternative approaches, such as PINNs, to overcome the shortcomings of traditional methods in modern applications. In this paper, an enhanced PINN methodology is introduced for the prediction of soliton structures for (2+ 1)-dimensional CNLSE \eqref{eq:3}. To the best of our knowledge, the application of PINNs to this class of equations has not been reported before. The proposed approach leverages a neuron-wise locally adaptive activation function to accelerate the convergence rate and improve the accuracy of standard PINN \cite{bib77}. By combining the proposed approach with the residual-based adaptive refinement strategy (RAR) \cite{bib78}, we demonstrate that each residual loss constructed from the model Eq. \eqref{eq:3} requires a different distribution of residual points within the computational domain. The performance and efficiency of the proposed framework will be evaluated in predicting various soliton dynamics through the lens of the $\mathbb{L}^2$ relative error. By considering different forms of dispersion and nonlinearity coefficients, vector parabolic dark one-soliton, vector m-shaped anti-dark one-soliton, and dark two-soliton are predicted. We also apply the proposed framework to the parameter discovery of the model Eq. \eqref{eq:3}. First, we apply the introduced PINN, equipped with a single neural network, for constant parameter discovery under different noise intensities. This PINN model not only predicts the solution of the model equation but also estimates the constant parameter as another trainable network parameter during training. Second, we employ a dual-network strategy for the proposed PINN scheme, where one neural network predicts the solution dynamics, while the other captures the behavior of the dispersion and nonlinearity variable coefficients. The obtained results indicate that the proposed framework is promising for exploring waveform propagation in the optical fiber communications landscape. The main contributions of this study are as follows:
\begin{itemize}
    \item Introduction of an enhanced PINN methodology for predicting soliton structures governed by the (2+ 1)-dimensional VC-CNLSE \eqref{eq:3} for the first time.
    \item Utilization of a neuron-wise locally adaptive activation function to accelerate convergence rate and improve the accuracy of standard PINN models.
    \item Integration of the residual-based adaptive refinement (RAR) strategy with the proposed PINN approach to allocate residual points optimally for different residual losses derived from the model equation.
    \item Estimation of constant parameters using a single neural network under various noise intensities.
    \item Discovery of variable coefficients using a dual-network strategy.
\end{itemize}

This is how the remainder of the paper is organized. In Sect.~\ref{sec:2}, we introduce the general framework of PINN algorithms for the general (2+ 1)-dimensional variable coefficient coupled nonlinear equations and discuss the related mechanisms. We also introduce our enhanced PINN framework in this section. We investigate the application of the proposed enhanced PINN algorithm for data-driven vector dark and anti-dark one- and two-soliton solutions to the (2+ 1)-dimensional VC-CNLSE. We illustratively show that our proposed algorithm accurately captures the solitonic dynamics. We also perform data-driven parameter discovery for the system of model equations to estimate the value of constant coefficients as well as the variable dispersion and nonlinearity coefficients. A conclusion is made in Sec.~\ref{sec:4}, where we summarize the effectiveness of the enhanced PINN framework and discuss its potential.
\section{PINNs for the (2+ 1)-dimensional VC-CNLSE}\label{sec:2}

The PINNs methodology leverages the approximation capability of deep neural networks to solve differential equations through a minimization problem. A feed-forward neural network with an appropriate architecture (in terms of depth and width) is trained to satisfy both the governing equations and the boundary and/or initial conditions by minimizing a loss function that induces physics-informed components and data-driven solutions. This approach allows the network to learn complex relationships within the data, ensuring that the solution adheres to the underlying physical laws while also capturing any observed patterns in the data.
In this section, we first introduce the general framework for PINNs methods applied to the (2+ 1)-dimensional coupled nonlinear equations with variable coefficients. Next, we introduce an enhanced PINN scheme and investigate the related modifications to the standard PINNs. In doing so, we consider the general (2+ 1)-dimensional variable coefficient coupled nonlinear equation in a complex environment space as follows:
\begin{equation} \label{eq:4}
i\psi_{j,t} + \mathcal{N}[\psi_{j}, \psi_{j,\bold{x}}, \psi_{j,\bold{xx}}, \cdots; \lambda] = 0,\quad \bold{x}\in\Omega = (\bold{x}_{L}, \bold{x}_{R}),\, t \in (T_{0}, T_{f}],
\end{equation}
subject to the following initial and boundary conditions
\begin{align}
\psi_{j}(\bold{x}, t = T_{0}) &= \psi_{j}^{ic}, \quad \bold{x} \in \Omega, \label{eq:5}\\
\psi_{j}(\bold{x}, t) &= \psi_{j}^{bc}, \quad (\bold{x}, t) \in \partial \Omega \times (T_{0}, T_{f}], \quad j = 1, 2. \label{eq:6}
\end{align}

In the problem setting above, $\psi_{j} = \psi_{j}(\bold{x}, t)$ represents a complex-valued function of the spatial variables $\bold{x} = (x, y)$ and the temporal variable $t$. $\mathcal{N}$ is a nonlinear operator acting on the solutions $\psi_{j}(\bold{x}, t)$ for $j = 1, 2$, their derivatives with respect to the spatial variables, and is parametrized by $\lambda$, which can be an unknown parameter (constant or variable). We also denote the domain of spatial variables by $\Omega = (\bold{x}_{L}, \bold{x}_{R})$ and the corresponding domain boundary by $\partial \Omega$. The computational interval for the temporal variable $t$ is also denoted by $[T_{0}, T_{f}]$. For $j = 1, 2$, physics models based on the left-hand side of Eq. \eqref{eq:4} can be formulated as:
\begin{equation}\label{eq:7}
\begin{cases}
\mathcal{F} := i\psi_{1,t} + \mathcal{N}[\psi_{1}, \psi_{1,\bold{x}}, \psi_{1,\bold{xx}}, \cdots; \lambda] = 0,\\
\mathcal{G} := i\psi_{2,t} + \mathcal{N}[\psi_{2}, \psi_{2,\bold{x}}, \psi_{2,\bold{xx}}, \cdots; \lambda] = 0.
\end{cases}
\end{equation}

Due to the complexity of the solutions $\psi_{j}(\bold{x}, t)$, we decompose them into the real and the imaginary parts as $\psi_{j}(\bold{x}, t) = u_{j}(\bold{x}, t) + i v_{j}(\bold{x}, t)$, where both $u_{j}(\bold{x}, t)$ and $v_{j}(\bold{x}, t)$ are real-valued functions. By considering the complex-valued physics models as $\mathcal{F} = \mathcal{F}_{u_{1}} + i\mathcal{F}_{v_{1}}$ and $\mathcal{G} = \mathcal{G}_{u_{2}} + i\mathcal{G}_{v_{2}}$, the system of equations \eqref{eq:7} can be reformulated as follows:
\begin{equation}\label{eq:8}
\begin{cases}
\mathcal{F}_{u_{1}} := u_{1,t} + \mathcal{N}_{u_{1}}[u_{1}, v_{1}, u_{2}, v_{2}, \cdots; \lambda] = 0,\\
\mathcal{F}_{v_{1}} := v_{1,t} + \mathcal{N}_{v_{1}}[u_{1}, v_{1}, u_{2}, v_{2}, \cdots; \lambda] = 0,\\
\mathcal{G}_{u_{2}} := u_{2,t} + \mathcal{N}_{u_{2}}[u_{1}, v_{1}, u_{2}, v_{2}, \cdots; \lambda] = 0,\\
\mathcal{G}_{v_{2}} := v_{2,t} + \mathcal{N}_{v_{2}}[u_{1}, v_{1}, u_{2}, v_{2}, \cdots; \lambda] = 0,
\end{cases}
\end{equation}
where $\mathcal{N}_{u_{1}}$, $\mathcal{N}_{v_{1}}$, $\mathcal{N}_{u_{2}}$, and $\mathcal{N}_{v_{2}}$ are specified nonlinear operators of the real-valued functions $u_{1}$, $v_{1}$, $u_{2}$, and $v_{2}$, respectively. A PINN model aims to predict these real-valued functions by minimizing a multi-objective loss function which is designed to minimize the discrepancy between the model's predictions to the initial and boundary conditions \eqref{eq:5} and \eqref{eq:6}, as well as the physical laws described by physics models in \eqref{eq:8}. The physics-informed part of the neural network model can be constructed using the provided physics models in \eqref{eq:8}, by differentiating the network’s outputs with respect to the spatial and temporal variables using the AD technique. Considering a three-dimensional input data $\bold{z}^{0} = (x, y, t)$ to the neural network model, a PINN model comprises a fully connected neural network with depth $D$, $D-1$ hidden layers, and an output layer. In each layer, we denote the number of neurons by $N^{\ell}$ for $\ell = 1, \cdots, D$. By feeding the input data into the network, an affine transformation will be applied as follows:
\begin{equation*}
\mathcal{L}^{\ell}(\bold{z}^{\ell-1}) := \bold{w}^{\ell}\bold{z}^{\ell-1} + \bold{b}^{\ell},\quad \ell = 1, \cdots, D,
\end{equation*}
where $\bold{z}^{\ell-1}$ is the input data from the previous layer to the next layer. Before sending the input data to the next layer, a nonlinear activation function $\sigma^{\ell}(\cdot)$ should be applied component-wise to the transformed data of each layer. This activation function introduces nonlinearity into the model, allowing the neural network to capture complex patterns and relationships within the data. In this paper, we use the tanh activation function for all our experiments. $\bold{w}^{\ell} \in \mathbb{R}^{N^{\ell}\times N^{\ell - 1}}$ and $\bold{b}^{\ell} \in \mathbb{R}^{N^\ell}$ are the weight matrix and bias vector corresponding to the layer $\ell$, respectively. So, the final output of the neural network can be formed as
\begin{equation*}
\hat{\psi}_{j,\boldsymbol{\theta}} := (\mathcal{L}^{D} \circ \sigma^{D-1}\circ \mathcal{L}^{D-1} \circ \cdots \circ \sigma^{1}\circ \mathcal{L}^{1})_{(\bold{z}^{0})},
\end{equation*}
where $\boldsymbol{\theta} := \{(\bold{w}^{\ell}, \bold{b}^{\ell})\}_{\ell = 1}^{D}$ is the set of all trainable parameters, and the operator $\circ$ is the composition operator. The PINNs model aims to predict the solutions to the system \eqref{eq:4}-\eqref{eq:6} by optimizing the network's trainable parameters. In doing so, we define the following overall loss function based on the mean square error (MSE) losses, considering the residual forms of Eq. \eqref{eq:5} and Eq. \eqref{eq:6}, along with the physics-informed component described by system \eqref{eq:8}
\begin{equation}\label{eq:9}
Loss(\boldsymbol{\theta}) := Loss_{ic}(\boldsymbol{\theta}) + Loss_{bc}(\boldsymbol{\theta}) + Loss_{r}(\boldsymbol{\theta}),
\end{equation}
where
\begin{align*}
Loss_{ic}(\boldsymbol{\theta}) &:= \frac{1}{N_{ic}} \sum_{k = 1}^{N_{ic}} w_{ic}^{k}\Big{(} \Big{\lvert} u_{1}^{k} - \hat{u}_{1, \boldsymbol{\theta}}(\bold{x}^{k}, t = T_{0}) \Big{\rvert}^2 +  \Big{\lvert} v_{1}^{k} - \hat{v}_{1, \boldsymbol{\theta}}(\bold{x}^{k}, t = T_{0}) \Big{\rvert}^2 \\
&+ \Big{\lvert} u_{2}^{k} - \hat{u}_{2, \boldsymbol{\theta}}(\bold{x}^{k}, t = T_{0}) \Big{\rvert}^2 +  \Big{\lvert} v_{2}^{k} - \hat{v}_{2, \boldsymbol{\theta}}(\bold{x}^{k}, t = T_{0}) \Big{\rvert}^{2}\Big{)}, \\
Loss_{bc}(\boldsymbol{\theta}) &:= \frac{1}{N_{bc}} \sum_{k = 1}^{N_{bc}}w_{bc}^{k} \Big{(} \Big{\lvert} u_{1}^{k} - \hat{u}_{1, \boldsymbol{\theta}}(\bold{x}^{k}, t^{k}) \Big{\rvert}^2 + \Big{\lvert} v_{1}^{k} - \hat{v}_{1, \boldsymbol{\theta}}(\bold{x}^{k}, t^{k}) \Big{\rvert}^{2}\\
&+ \Big{\lvert} u_{2}^{k} - \hat{u}_{2, \boldsymbol{\theta}}(\bold{x}^{k}, t^{k}) \Big{\rvert}^2 + \Big{\lvert} v_{2}^{k} - \hat{v}_{2, \boldsymbol{\theta}}(\bold{x}^{k}, t^{k}) \Big{\rvert}^2 \Big{)},
\end{align*}
and
\begin{align*}
Loss_{r}(\boldsymbol{\theta}) &:= \frac{1}{N_{r}} \sum_{k = 1}^{N_{r}} w_{r}^{k}\Big{(} \Big{\lvert} \mathcal{F}_{\hat{u}_{1,\boldsymbol{\theta}}}(\bold{x}^{k}, t^{k}) \Big{\rvert}^{2} + \Big{\lvert} \mathcal{F}_{\hat{v}_{1,\boldsymbol{\theta}}}(\bold{x}^{k}, t^{k}) \Big{\rvert}^{2} \\
&+ \Big{\lvert} \mathcal{G}_{\hat{u}_{2,\boldsymbol{\theta}}}(\bold{x}^{k}, t^{k}) \Big{\rvert}^{2} + \Big{\lvert} \mathcal{G}_{\hat{v}_{2,\boldsymbol{\theta}}}(\bold{x}^{k}, t^{k}) \Big{\rvert}^{2} \Big{)},
\end{align*}
where $\Big{\{}(\bold{x}^{k}, t = T_{0}), u_{1}^{k}, v_{1}^{k}, u_{2}^{k}, v_{2}^{k}\Big{\}}_{k = 1}^{N_{ic}}$ and $\Big{\{}(\bold{x}^{k}, t^{k}), u_{1}^{k}, v_{1}^{k}, u_{2}^{k}, v_{2}^{k}\Big{\}}_{k = 1}^{N_{bc}}$ are the sets of training data corresponding to the initial and boundary conditions, respectively. We also denote the set of residual points by $\Big{\{} (\bold{x}^{k}, t^{k})\Big{\}}_{k = 1}^{N_{r}}$ for encoding the physical laws $\mathcal{F}_{\hat{u}_{1,\boldsymbol{\theta}}}$, $\mathcal{F}_{\hat{v}_{1,\boldsymbol{\theta}}}$, $\mathcal{G}_{\hat{u}_{2,\boldsymbol{\theta}}}$, and $\mathcal{G}_{\hat{v}_{2,\boldsymbol{\theta}}}$ during training phase, constructed in \eqref{eq:8}. These points can be sampled either uniformly or randomly within the computational domain. $w_{ic}$, $w_{bc}$, and $w_{r}$ are point-wise loss weights that specify different weights assigned to the initial condition, boundary condition, and residual loss terms in the overall loss function, respectively. These weights allow for fine-tuning the relative importance of each component, enabling the model to balance the accuracy of the predictions across the different aspects of the problem. Then, the optimal values of the network's parameters can be obtained by minimizing the overall loss function \eqref{eq:9} using variants of the gradient descent algorithm (e.g., SGD and Adam) \cite{bib80,bib81}, or other classes of optimization algorithms. The corresponding complex-valued physics-informed part of the PINN model for the (2+ 1)-dimensional VC-CNLSE \eqref{eq:3} can be formulated as follows:
\begin{equation}
\begin{cases} \label{eq:10}
\mathcal{F}_{u_{1}} := -v_{1,t} + \alpha(t)(u_{1,xx} + u_{1,yy}) + \beta(t)(u_{1}^2 + v_{1}^2 + u_{2}^2 + v_{2}^2)u_{1}= 0,\\
\mathcal{F}_{v_{1}} := u_{1,t} + \alpha(t)(v_{1,xx} + v_{1,yy}) + \beta(t)(u_{1}^2 + v_{1}^2 + u_{2}^2 + v_{2}^2)v_{1}= 0,\\
\mathcal{G}_{u_{2}} := -v_{2,t} + \alpha(t)(u_{2,xx} + u_{2,yy}) + \beta(t)(u_{1}^2 + v_{1}^2 + u_{2}^2 + v_{2}^2)u_{2}= 0,\\
\mathcal{G}_{v_{2}} := u_{2,t} + \alpha(t)(v_{2,xx} + v_{2,yy}) + \beta(t)(u_{1}^2 + v_{1}^2 + u_{2}^2 + v_{2}^2)v_{2}= 0.
\end{cases}
\end{equation}

A major drawback of neural network-based approaches is their slow training, which can affect the convergence speed and performance of the underlying model. To address this issue, a locally adaptive activation function is introduced \cite{bib77}, in which scalable parameters are defined to increase the slope of the activation functions. This further reduces the risk of vanishing gradients and improves the convergence speed, resulting in lower training costs. For a neural network model parametrized by $\boldsymbol{\theta}$, the neuron-wise scalable parameters $na$ can act as
\begin{equation*}
\sigma^{\ell}(na_{k}^{\ell}(\mathcal{L}^{\ell}(z^{\ell-1}))_{k}), \quad \ell = 1, \cdots, D-1,\,\,\, k = 1, \cdots, N^{\ell},
\end{equation*}
where $n\geq 1$ is a pre-defined scalable factor. This configuration will introduce additional $\sum_{\ell = 1}^{D-1} N^{\ell}$ trainable parameters to be optimized, where each neuron in each hidden layer has its own slope for the activation function. Then, the set of trainable parameters $\boldsymbol{\theta}$ consists of the set of weights and biases $\{(\bold{w}^{\ell}, \bold{b}^{\ell})\}_{\ell = 1}^{D}$ and the set of scalable parameters $\{a_{k}^{\ell}\}_{\ell = 1}^{D-1}$, $\forall k = 1,\cdots,N^{\ell}$. We initialize these scalable parameters such that $na_{k}^{\ell} = 1, \,\,\, \forall n\geq 1$. In order to force the network to quickly increase the slope of the activation function, the authors in \cite{bib77} defined a slope recovery loss term to be added into the overall loss function as:
\begin{equation*}
Loss_{a}(\boldsymbol{\theta}) := \frac{1}{\frac{1}{D-1}\sum_{\ell = 1}^{D-1}\exp(\frac{\sum_{k = 1}^{N^{\ell}}a_{k}^{\ell}}{N^{\ell}})}.
\end{equation*}

Hence, the loss function \eqref{eq:9} can be reformulated as follows:
\begin{equation} \label{eq:11}
Loss(\boldsymbol{\theta}) := Loss_{ic}(\boldsymbol{\theta}) + Loss_{bc}(\boldsymbol{\theta}) + Loss_{r}(\boldsymbol{\theta}) + w_{a}\cdot Loss_{a}(\boldsymbol{\theta}),
\end{equation}
where $w_{a}$ is a user-defined hyper-parameter weight for the individual loss function corresponding to the slope recovery term. 

In applying the proposed PINN framework for data-driven solution and parameter discovery to the model equation \eqref{eq:3}, the loss function \eqref{eq:11} ensures that the model predictions satisfy the observed data corresponding to the initial and boundary conditions, as well as the physics-informed parts provided by the system of equations \eqref{eq:10}. In general, for a PINN model, the behavior of the equation residuals often exhibits many oscillations around zero. Indeed, the underlying PINN model aims to force the equation residuals to zero by penalizing these residuals at so-called residual points. The residual-based adaptive refinement (RAR) strategy is proposed to address this issue. The RAR strategy focuses on regions within the computational domain where the residuals are large. This approach improves the distribution of residual points by adaptively sampling more points in these regions during the training phase. In all our experiments, we randomly sample m new residual points after specific training iterations until either the maximum number of iterations is reached or the mean equation residuals fall below a pre-defined threshold. This motivates us to define region-specific weights for individual residual loss terms in Eq. \eqref{eq:11}, corresponding to the physics-informed parts described by the governing equations. Indeed, the loss term corresponding to the physics-informed parts can be reformulated as follows:
\begin{align*}
Loss_{r}(\boldsymbol{\theta}) &:= \frac{1}{N_{r}} \sum_{k = 1}^{N_{r}}\Big{(} w_{r}^{\mathcal{F}_{\hat{u}}}\Big{\lvert} \mathcal{F}_{\hat{u}_{1,\boldsymbol{\theta}}}(\bold{x}^{k}, t^{k}) \Big{\rvert}^{2} + w_{r}^{\mathcal{F}_{\hat{v}}}\Big{\lvert} \mathcal{F}_{\hat{v}_{1,\boldsymbol{\theta}}}(\bold{x}^{k}, t^{k}) \Big{\rvert}^{2} \\
&+ w_{r}^{\mathcal{G}_{\hat{u}}}\Big{\lvert} \mathcal{G}_{\hat{u}_{2,\boldsymbol{\theta}}}(\bold{x}^{k}, t^{k}) \Big{\rvert}^{2} + w_{r}^{\mathcal{G}_{\hat{v}}}\Big{\lvert} \mathcal{G}_{\hat{v}_{2,\boldsymbol{\theta}}}(\bold{x}^{k}, t^{k}) \Big{\rvert}^{2} \Big{)},
\end{align*}
where $w_{r}^{\mathcal{F}_{\hat{u}}}, w_{r}^{\mathcal{F}_{\hat{v}}}, w_{r}^{\mathcal{G}_{\hat{u}}},$ and $w_{r}^{\mathcal{G}_{\hat{v}}}$ are region-specific weights to specify different importance of each residual equation during training. Indeed, the value of these weights can be regarded as the sampled residual points within the computational domain using the RAR strategy during training. The proposed enhanced PINN framework introduces a region-specific sampling strategy for residual equations, allowing for different distributions of residual points tailored to each residual equation. In the standard PINN framework, a large and often uniformly distributed set of residual points is used throughout training. Since PINNs rely on automatic differentiation to compute derivatives of network outputs with respect to input variables, increasing the number of residual points significantly adds to the computational complexity. Consequently, training a standard PINN with many residual points can be computationally expensive and time-consuming. In contrast, the enhanced PINN’s targeted sampling strategy reduces the total number of residual points required, thereby lowering the training time. Moreover, by customizing the distribution of residual points for each residual equation, the enhanced PINN achieves improved convergence speed and overall training efficiency.

\begin{figure}[h!]
\centering
\includegraphics[height = 6.5cm]{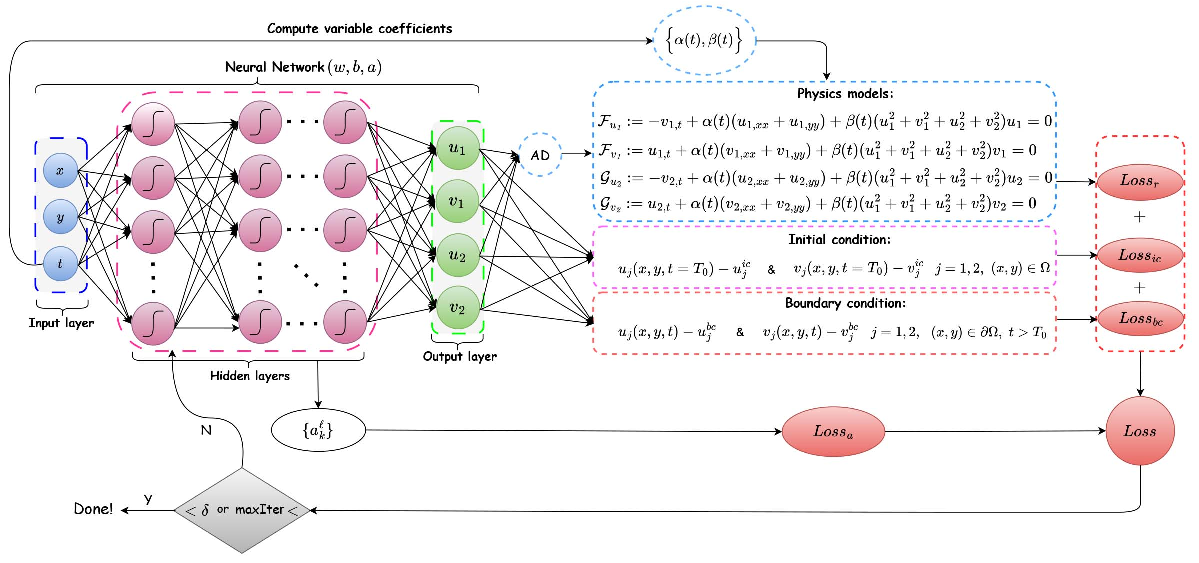}
\caption{The architecture of the proposed PINN for the (2+ 1)-dimensional VC-CNLSE.}
\label{fig:1}
\end{figure}

The architecture of the proposed PINN framework is depicted in Fig.~\ref{fig:1}, including an input layer for spatial and temporal coordinates, inducing a three-dimensional problem solving. The input data is transformed by applying the weights and biases to each neuron in each hidden layer. An adaptive nonlinear activation function is then applied component-wise to the transformed data, activating each neuron. Each neuron in the previous layer is fully connected to the neurons in the next layer. In the output layer, there are four neurons, corresponding to the predictions of the system \eqref{eq:3}, which include the real and imaginary parts of the decomposed solutions. After the network's outputs are obtained, the network is able to construct the physics-informed parts through the application of the AD technique. For this, the values of the variable coefficients must be computed first. Then, the physics-informed neural network can be constructed using the residuals provided by the physics models in \eqref{eq:10}. Both networks share their parameters during training. By constructing the loss functions corresponding to the governing equations, initial and boundary conditions, and the slope recovery loss, one can repeatedly obtain the optimal values of the network's parameters by minimizing the overall loss function using appropriate optimization algorithms. In this paper, the loss functions are optimized using both the Adaptive Moment Estimation (Adam) and Limited-Memory Broyden-Fletcher-Goldfarb-Shanno (L-BFGS) algorithms. Adam is a gradient-based optimization algorithm that combines the advantages of two other extensions of the stochastic gradient descent algorithm. On the other hand, L-BFGS is a full-batch quasi-Newton method known for its effectiveness in optimizing functions by approximating the inverse Hessian matrix \cite{bib82}. Combining these two methods leverages Adam’s robustness and adaptability in the initial stages of training with L-BFGS’s precision in fine-tuning the final solution. This minimization process will continue until the network's error falls below a pre-defined threshold $\delta$ or the maximum number of iterations is reached.
We sketch the proposed PINN algorithm for solving the (2+ 1)-dimensional VC-CNLSE in Algorithm~\ref{IPINN}. 
\begin{algorithm} 
\caption{Enhanced PINN algorithm to (2+ 1)-dimensional VC-CNLSE}\label{IPINN}
\begin{algorithmic}

\Require
\\
\begin{description}
\item[$\bullet$] Training data sets $\Big{\{}(\bold{x}^{k},T_{0}), u_{1}^{k}, v_{1}^{k}, u_{2}^{k}, v_{2}^{k}\Big{\}}_{k = 1}^{N_{ic}}$ and 

$\Big{\{}(\bold{x}^{k}, t^{k}), u_{1}^{k}, v_{1}^{k}, u_{2}^{k}, v_{2}^{k}\Big{\}}_{k = 1}^{N_{bc}}$
\item[$\bullet$] Residual points $\Big{\{} (\bold{x}^{k}, t^{k})\Big{\}}_{k = 1}^{N_{r}}$
\item[$\bullet$] The value of the hyper-parameters $\delta$, $m$, $n$, $w_{a}$
\item[$\bullet$] Construct a neural network model with Xavier initialization technique
\end{description}
\While { (\text{error} $>$ $\delta$ \, \text{and}\, \text{nIter}$<$\text{maxIter})}

(a) Construct the physics-informed neural network by computing the variable coefficients and incorporating the network's predictions into the physics models using the AD technique.

(b) Specification of the overall loss function $Loss(\boldsymbol{\theta})$ using the losses corresponding to the governing equations, initial and boundary conditions, and the slope recovery loss function.

(c) Add $m$ new residual points for each residual equation based on the RAR technique, where 
\begin{equation*}
\argmaxA_{(\bold{x}, t)} \Big{\{} \lvert \mathcal{F}_{u_{1}} \rvert, \lvert \mathcal{F}_{v_{1}} \rvert, \lvert \mathcal{G}_{u_{2}} \rvert, \lvert \mathcal{G}_{v_{2}} \rvert \Big{\}}
\end{equation*}

(d) Find the optimal values of the network's parameters by minimizing the overall loss function using an appropriate optimization algorithm.
\EndWhile
\end{algorithmic}
\end{algorithm}

\section{Results and discussions}\label{sec:3}

In this section, we apply the proposed PINN framework to the (2+ 1)-dimensional VC-CNLSE and assess its accuracy and performance in predicting various soliton dynamics. First, we address the problem of data-driven soliton solutions to the model equation, predicting phenomena such as vector parabolic dark one-soliton, vector m-shaped anti-dark one-soliton, and dark two-solitons. By considering different formations of dispersion and nonlinearity coefficients, we demonstrate that our method is capable of predicting the corresponding soliton dynamics with high accuracy. Second, we explore the parameter discovery within the model equation, focusing on accurately identifying both constant and variable coefficients. In all our experiments, we initialize the scalable parameters such that $n = 5$ and $a_{k}^{\ell} = 0.2$ for $\ell = 1, \cdots, D-1$ and $k = 1, \cdots, N^{\ell}$ and initialize the network's parameters using the Xavier initialization technique. We set the initial condition loss, boundary condition loss, and slope recovery loss weights as: $w_{ic} = 1$, $w_{bc} = 1$, and $w_{a} = \frac{1}{100}$. Setting $w_{a} = \frac{1}{100}$ ensures that the value of the loss is not too large. We generate the residual points randomly using the Latin Hypercube Sampling (LHS) strategy \cite{bib83} within the computational domain. We start training the constructed PINN model with an initial number of residual points, setting $m = 25$. After every $5000$ iterations, we add $2^{k}\times m$ new residual points for each residual equation based on the RAR algorithm during the training phase, where $k$ denotes the number of times the $5000$-iteration update has occurred, and this sampling process is continued until either the maximum number of iterations is reached or the mean equation residuals fall below $1e-6$. We set the stopping criterion to $\delta = 1e-05$ during the training of the underlying PINN model. The accompanying code for this paper is written in Python, and the numerical results are reported using the JAX library. The JAX library is used for its efficient automatic differentiation and high-performance numerical computing capabilities. The computations were performed on a Lenovo Legion laptop equipped with a 2.30 GHz Intel Core i7-11800H processor and a single NVIDIA GeForce RTX 3060 GPU card. To evaluate the performance and accuracy of our method, we consider two metrics, the $\mathbb{L}^{2}$ error and the point-wise error, defined as follows:
\begin{align*}
Error_{\mathbb{L}^2} &:= \frac{\Big{(} \sum_{k = 1}^{N} \Big{\lvert} \psi_{j}(\bold{x}^{k}, t^{k}) - \hat{\psi}_{j, \boldsymbol{\theta}}(\bold{x}^{k}, t^{k}) \Big{\rvert}^{2} \Big{)}^{1/2}}{\Big{(} \sum_{k = 1}^{N} \Big{\lvert} \psi_{j}(\bold{x}^{k}, t^{k}) \Big{\rvert}^{2} \Big{)}^{1/2}},\\
Error_{pw} &:=\frac{\Big{\lvert} \psi_{j}(\bold{x}^{k}, t^{k}) - \hat{\psi}_{j, \boldsymbol{\theta}}(\bold{x}^{k}, t^{k}) \Big{\rvert}}{\Big{\lvert} \psi_{j}(\bold{x}^{k}, t^{k}) \Big{\rvert}}, \quad j = 1, 2,
\end{align*}
where $\psi_{j}(\cdot, \cdot)$ and $\hat{\psi}_{j, \boldsymbol{\theta}}(\cdot, \cdot)$ are the exact and predicted solutions, respectively, evaluated at sampled test points.
\subsection{Data-driven solution to (2+ 1)-dimensional VC-CNLSE}

Choosing different and appropriate formations of dispersion and nonlinearity coefficients leads us to explore a wide range of solitonic dynamics. By adjusting these coefficients, we can model various physical scenarios. Here, we aim to employ the proposed PINNs framework based on Algorithm~\ref{IPINN}, considering the equation system \eqref{eq:3} equipped with the initial condition \eqref{eq:5} and the boundary condition \eqref{eq:6} for data-driven solution of the (2+ 1)-dimensional VC-CNLSE. For each experiment, the network configurations for the underlying PINN model and the hyper-parameter settings are reported in Table~\ref{tab:1}.
\begin{table}[h!]
    \centering
    \caption{Network configurations and hyper-parameter settings for experimental cases.}
    \begin{tabular}{ccccccc}
    \toprule
    &&&&&Network&\\
    \cmidrule(r){6-7} 
    Cases&$N_{ic}$&$N_{bc}$ &Iterations&Learning rate&Depth &Width\\
    \midrule
    1&$100$&$100$&$35,144$&$0.001$&$5$ &$40$ \\
    2&$100$&$100$&$35,170$&$0.001$&$5$&$40$ \\
    3&$200$&$200$&$35,215$&$0.0001$&$4$ &$100$\\
    4&$200$&$200$&$35,034$&$0.0001$&$4$&$100$\\
    \bottomrule
    \end{tabular}
    \label{tab1}
\end{table}

\subsection{Discovery of one-soliton solutions}

By applying the Hirota method, one can obtain the following one-soliton solutions \cite{bib25}:
\begin{equation}\label{eq:12}
\psi_{1}(x, y, t) = \mu\frac{e^{ib(t)(1 + e^{\xi(x,y,t) + 2i\theta})}}{1+e^{\xi(x,y,t)}},\quad \psi_{2}(x, y, t) = \lambda\frac{e^{ib(t)(1 + e^{\xi(x,y,t) + 2i\theta})}}{1+e^{\xi(x,y,t)}},
\end{equation}
where 
\begin{align}
b(t) &:= (\lvert \lambda \rvert^2 + \lvert \mu \rvert^2)\int \beta(t) dt,\quad \beta(t) := -\frac{\csc^{2}(\theta)(k^2 + m^2)\alpha(t)}{2(\lvert \lambda \rvert^2 + \lvert \mu \rvert^2)}\label{eq:13},\\
w(t) &:= \cot(\theta)(k^2 + m^2)\int \alpha(t) dt, \quad
\xi(x,y,t) := kx + my + w(t) + \epsilon. \label{eq:16}
\end{align}

Here, $k$, $m$, and $\epsilon$ are real constants. $\lambda$ and $\mu$ are complex constants, while $\theta$ is a real phase shift. One can see that the one-soliton solutions \eqref{eq:12} are proportional to each other. This leads us to dealing with a Manakov system when we set $\tau = 1$ in the system of equations \eqref{eq:3}.

\noindent\textbf{Case 1:} Parabolic vector dark one-soliton
\label{C:1}

\begin{figure}[h!]
\centering
\includegraphics[height = 12cm,width = 13cm]{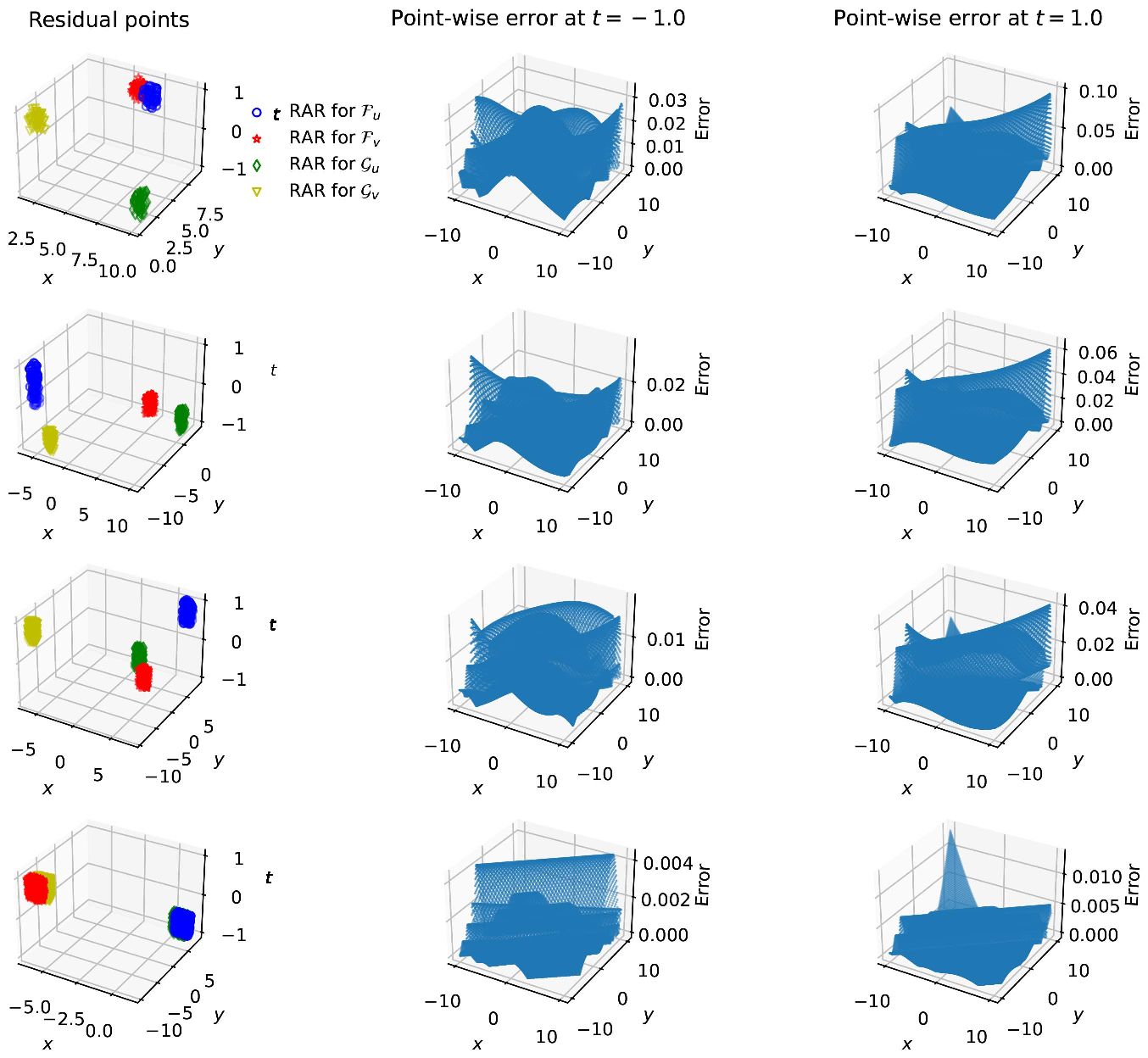}
\caption{The distribution of added new residual points (first column) and the point-wise errors of $\psi_{1}$ (second and third columns).}
\label{fig:2}
\end{figure}

\begin{figure}[h!]
\centering
\includegraphics[height = 9cm,width = 13cm]{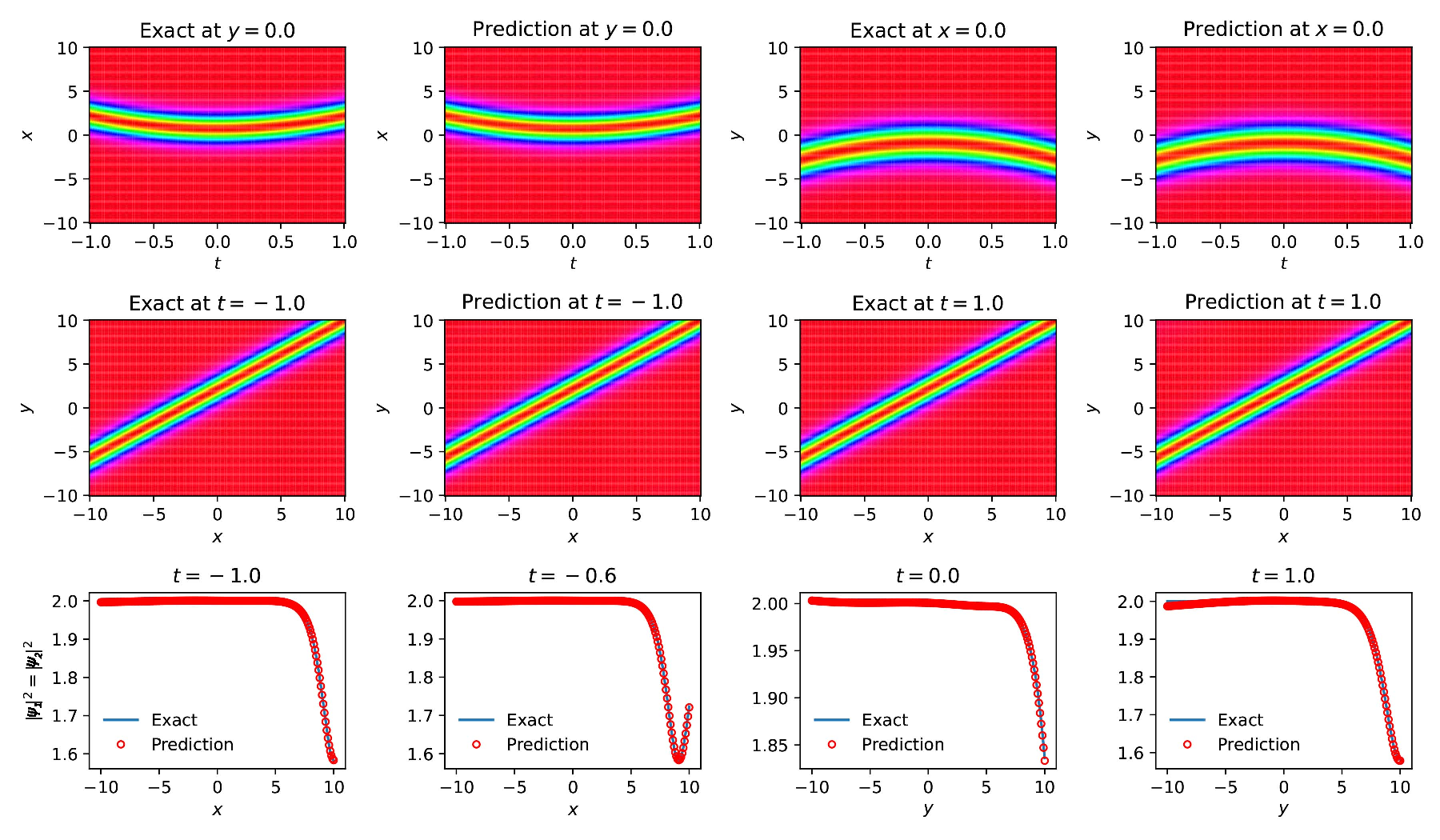}
\caption{The density plots of the exact and predicted dynamics (top and middle rows) and the comparison between the exact and predicted solutions at different temporal levels (bottom row).}
\label{fig:3}
\end{figure}

\begin{figure}[h!]
\centering
\includegraphics[height = 7cm,width = 13cm]{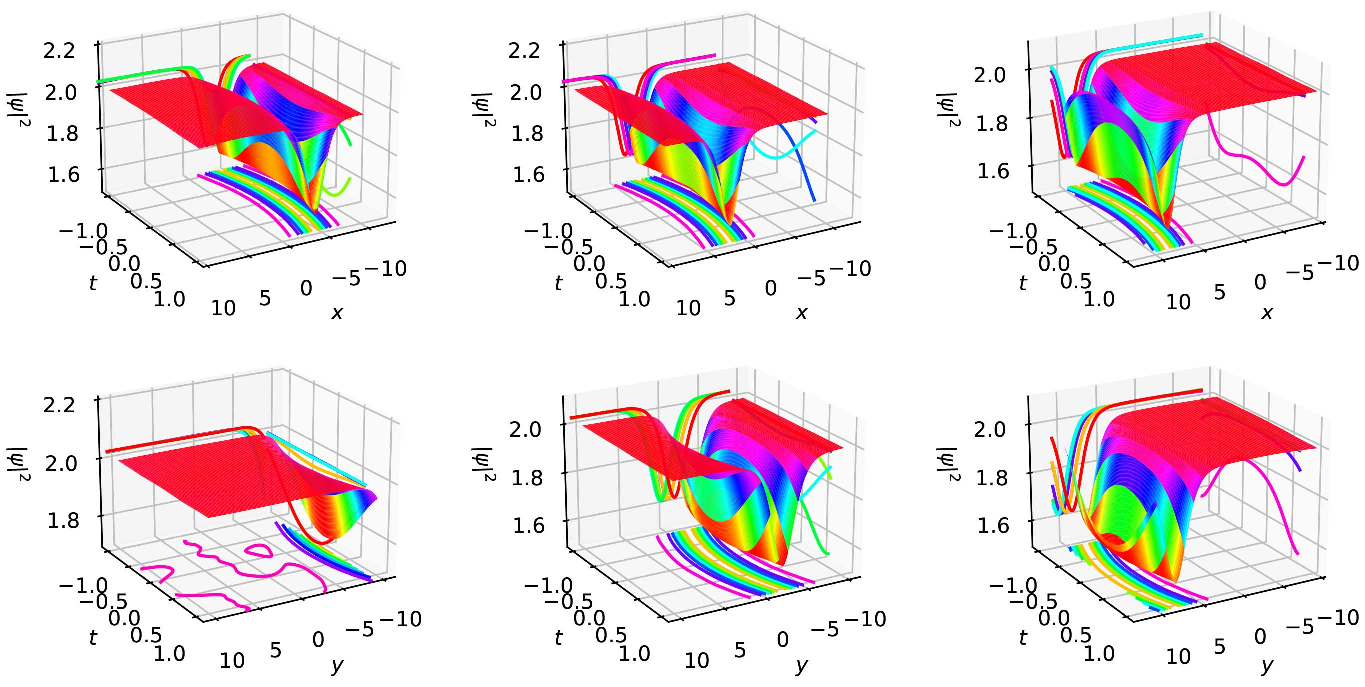}
\caption{Three-dimensional plots with contour map of the predicted dynamics for $\psi_{j}$ in the $x-t$ (top row) and $y-t$ (bottom row) planes.}
\label{fig:4}
\end{figure}

\begin{figure}[h!]
\centering
\includegraphics[height = 3.5cm,width = 13.5cm]{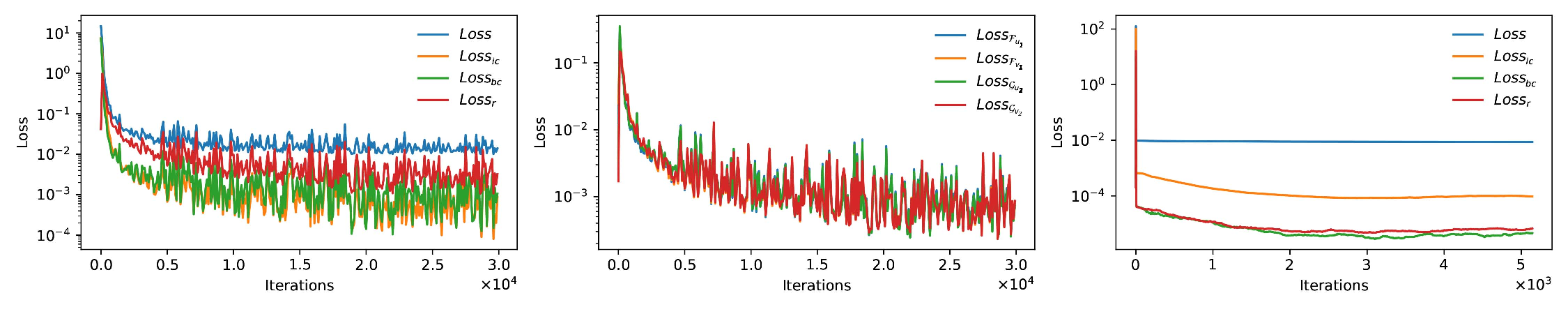}
\caption{The loss curves during the training with $30,000$ iterations of the Adam optimizer (left and middle panels) and the loss curves with $5,144$ iterations of the L-BFGS optimizer (right panel).}
\label{fig:5}
\end{figure}

Based on solutions \eqref{eq:12}, a parabolic vector dark one-soliton for the (2+ 1)-dimensional VC-CNLSE \eqref{eq:3} is obtained with parameters $k=-1.4$, $m=1.1$, $\epsilon=1$, $\mu=-2$, $\theta=3.8$, $\lambda=-2$, and $\alpha(t)=t$, yielding equal amplitudes for $\psi_{1}$ and $\psi_{2}$. We set the spatial domain $\Omega = [-10, 10]\times [-10, 10]$ and the temporal domain $[-1, 1]$. We divide the computational domain $\Omega \times [-1, 1]$ into $256 \times 256 \times 201$ discrete equidistant points to randomly generate $N_{ic} = N_{bc} = 100$ initial and boundary training points based on the information above. We use the LHS technique to generate $N_{r} = 10,000$ residual points in the first stage of training and then gradually add new residual points based on Algorithm~\ref{IPINN}. For this problem, we consider a neural network model with $5$ hidden layers and $40$ neurons in each layer. The network parameters are optimized using $30,000$ iterations of the Adam optimizer followed by $5,144$ iterations of the L-BFGS optimizer, reducing the $\mathbb{L}^2$ relative errors for $\psi_{1}$ and $\psi_{2}$ from $0.18\%$ and $0.19\%$ to $0.073\%$ and $0.064\%$, respectively. As depicted in the first column of Fig.~\ref{fig:2}, the added residual points corresponding to the residual equations have different distributions, demonstrating that each residual equation exhibits distinct characteristics, highlighting the need to consider their individual importance when applying PINN methods. The three-dimensional plots of point-wise errors for the solution $\psi_{1}$ along the $x-y$ axis at two temporal levels, $t = -1$ and $t = 1$, are depicted in the second and third columns of Fig.~\ref{fig:2}, respectively. One can see that incorporating RAR significantly reduces prediction errors and yields a more uniform distribution of residual points by driving the residuals towards zero during training. In Fig.~\ref{fig:3}, the density plots of the exact and predicted dynamics, as well as the comparison between the exact and predicted solutions for $\psi_{1}$ at different temporal levels, are depicted. In the top row, the density plots along the $x-t$ plane at $y = 0$ and the $y-t$ plane at $x = 0$ are shown, where the pulse propagates parabolically in opposite directions in the $x-t$ and $y-t$ planes. In the middle row, the density plots of the exact and predicted dynamics at two temporal levels, $t = -1$ and $t = 1$, along the $x-y$ plane are depicted. One can observe that the pulse propagates along a straight line. A comparison between the exact and predicted solutions at four temporal levels for $x = 10$ and $y =10$ is presented in the bottom row. These results show that the trained PINN model can accurately capture the exact dynamics in the $x-t$, $y-t$, and $x-y$ planes, demonstrating the robustness of our method. The evolution of the parabolic dark one-soliton in the $x-t$ plane (first row) at $y = -8$, $y = 0$, and $y = 8$ (from left to right) and in the $y-t$ plane (bottom row) at $x = -8$, $x = 0$, and $x = 8$ (from left to right) is illustrated in Fig.~\ref{fig:4}. The loss trajectories associated with the Adam optimizer and the L-BFGS optimizer are depicted in Fig.~\ref{fig:5}. From the left panel, one can see that the Adam optimizer exhibits a slowly decreasing loss with several oscillations during training. From the middle panel of Fig.~\ref{fig:5}, one can see that the oscillatory behavior of the MSE loss for each individual residual equation is relatively close to that of the others. In contrast, the L-BFGS optimizer exhibits smoother behavior of the loss curves without any oscillations as depicted in the right panel of Fig.~\ref{fig:5}. It should be noted that, in comparison with the results obtained from the Adam optimizer, the L-BFGS optimizer demonstrates a faster convergence framework. Table \ref{tab2} compares the performance of the proposed enhanced PINN method, the adaptive activation function-based PINN method in \cite{bib77}, and the standard PINN method in terms of $\mathbb{L}^{2}$ relative error, loss values (Adam followed by L-BFGS), and training time (in seconds). For a fair comparison, all networks are configured with $5$ hidden layers and $40$ neurons per layer. For the method in \cite{bib77}, the adaptive activation functions are set the same as in the enhanced PINN. The number of residual points is set to $N_{r} = 16,200$ for both the method in \cite{bib77} and the standard PINN. These results show that the proposed enhanced PINN method outperforms both the PINN method in \cite{bib77} and the standard PINN method by achieving the lowest $\mathbb{L}^{2}$ relative errors and loss values, along with a shorter training time. This improvement is due to its region-specific training approach, which focuses on areas with high equation residuals and allocates different distributions of residual points for each residual equation, resulting in more effective penalization of the equation residuals during training while using a smaller number of residual points. In contrast, the other methods use large datasets with the same distribution for all residual equations, leading to higher computational cost and longer training times due to the use of the AD technique, which requires computing the derivatives of the network outputs with respect to input data at all residual points during training. Overall, the enhanced PINN demonstrates superior accuracy and computational efficiency for solving the (2+ 1)-dimensional VC-CNLSE problem. 
\begin{table}[h!]
    \centering
    \caption{Comparison results in terms of the $\mathbb{L}^{2}$ relative error, loss value, and training time (s).}
    \begin{tabular}{cccccc}
    \toprule
    &$\mathbb{L}^{2}$ error&&Loss value&&\\
    \cmidrule(r){2-3} \cmidrule(r){4-5} 
    Method&$\psi_{1}$&$\psi_{2}$ &Adam&L-BFGS&Training time (s)\\
    \midrule
    Enhanced PINN&$0.073\%$&$0.064\%$&$1.3371e-02$&$8.4830e-03$&$2510.1597$\\
    Method in \cite{bib77}&$0.37\%$&$0.55\%$&$9.4780e-01$&$8.1209e-02$&$2741.0294$\\
    PINN&$5.12\%$&$4.28\%$&$3.0813e-00$&$6.3085e-01$&$2571.8029$\\
    \bottomrule
    \end{tabular}
    \label{tab2}
\end{table}

\noindent\textbf{Case 2:} Vector m-shaped anti-dark one-soliton \label{C:2}

\begin{figure}[h!]
\centering
\includegraphics[height = 9cm,width = 13cm]{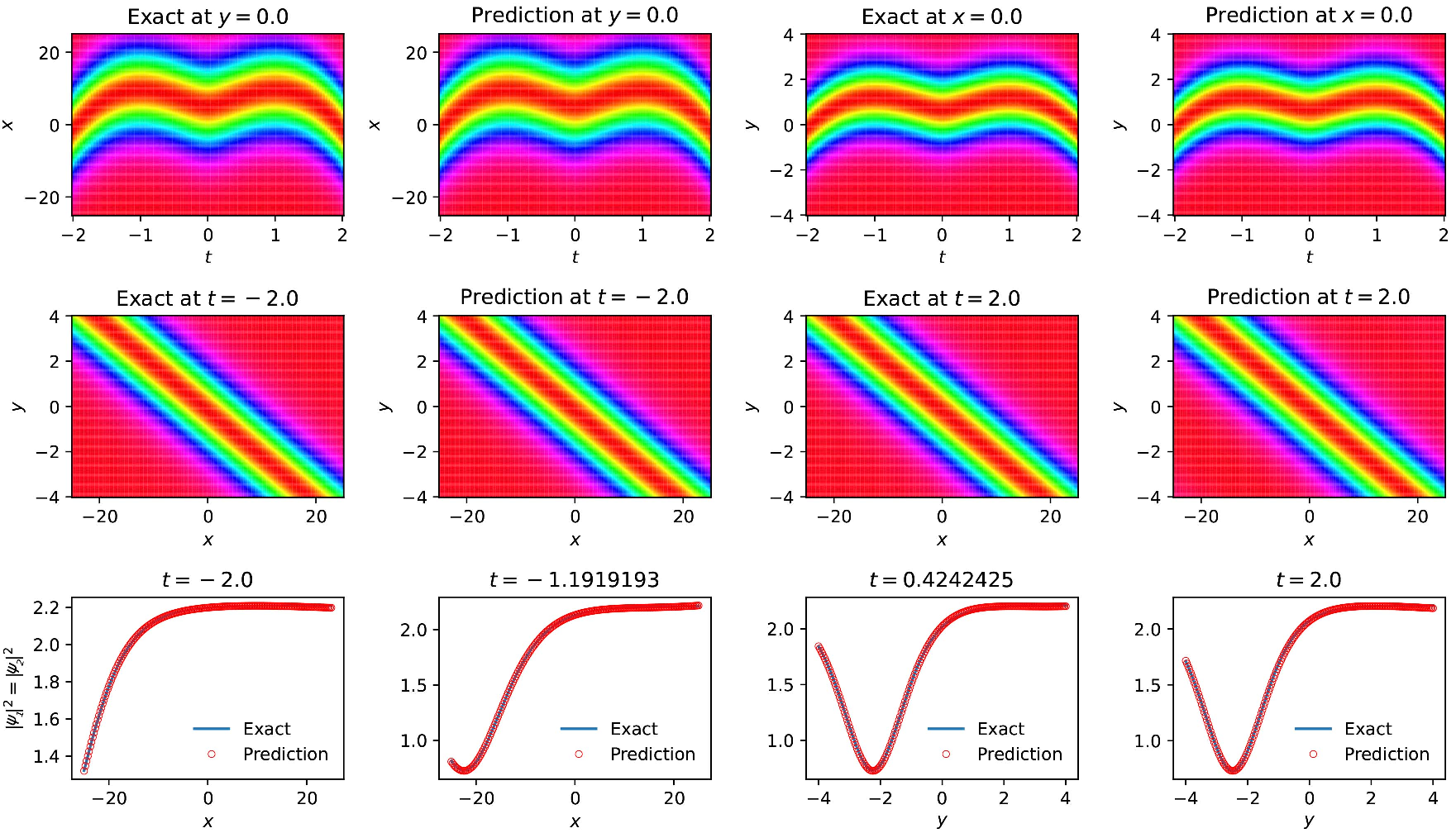}
\caption{The density plots of the exact and predicted dynamics (top and middle rows) and the comparison between the exact and predicted solutions at different temporal levels (bottom row).}
\label{fig:6}
\end{figure}

\begin{figure}[h!]
\centering
\includegraphics[height = 3cm,width = 13cm]{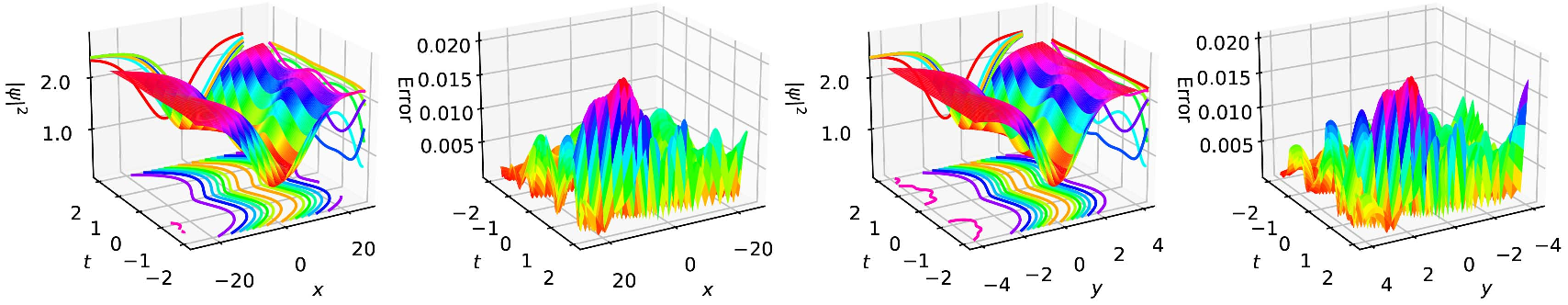}
\caption{Three-dimensional plots with contour maps and the corresponding point-wise errors for $\psi_{j}$ in the $x-t$ plane (first and second panels) and $y-t$ plane (third and fourth panels), from left to right.}
\label{fig:7}
\end{figure}

\begin{figure}[h!]
\centering
\includegraphics[height = 3.5cm,width = 13.5cm]{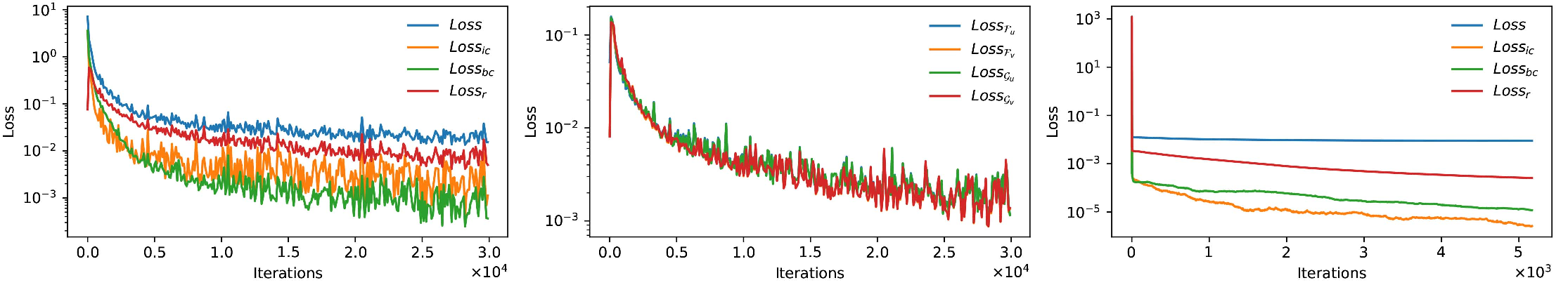}
\caption{The loss curves during the training with $30,000$ iterations of the Adam optimizer (left and middle panels) and the loss curves with $5,170$ iterations of the L-BFGS optimizer (right panel).}
\label{fig:8}
\end{figure}

\begin{figure}[h!]
\centering
\includegraphics[height = 4.5cm,width = 13.5cm]{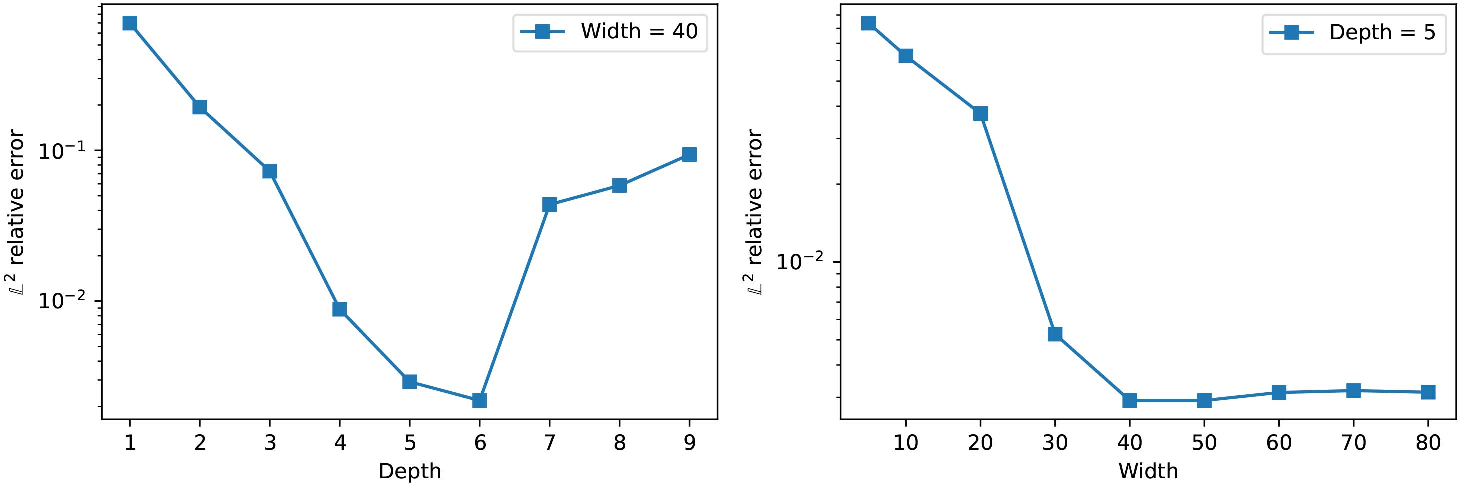}
\caption{The effect of the depth and width of the network on the $\mathbb{L}^{2}$ relative error of $\psi_{1}$.}
\label{fig:88}
\end{figure}

\begin{figure}[h!]
\centering
\includegraphics[height = 3.5cm,width = 13.5cm]{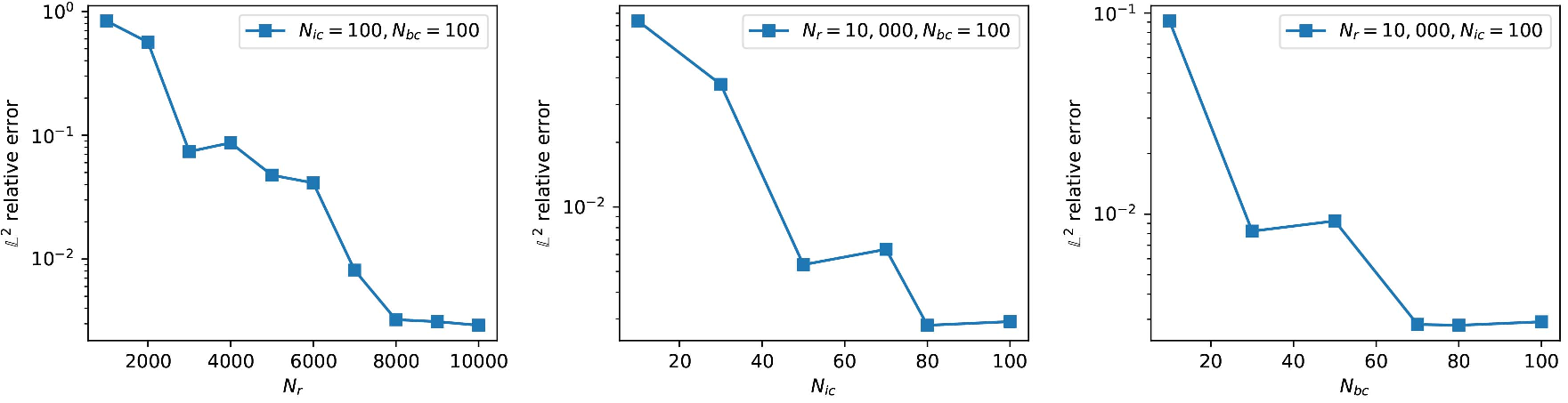}
\caption{The effect of the number of training and residual points on the $\mathbb{L}^{2}$ relative error of $\psi_{1}$.}
\label{fig:888}
\end{figure}

By setting the parameters as $k = -0.19$, $m = -1.5$, $\epsilon = 1$, $\mu = -1.1 + i$, $\theta = -4.1$, $\lambda = -1.1 + i$, and $\alpha(t) = t \ln t^2$ in equations \eqref{eq:12}-\eqref{eq:16}, a vector m-shaped anti-dark one-soliton solution to the (2+ 1)-dimensional VC-CNLSE is obtained, where both $\psi_{1}$ and $\psi_{2}$ have the same amplitudes. The spatial domain is $\Omega = [-25, 25] \times [-4, 4]$ and the temporal domain is $[-2, 2]$, discretized into $256 \times 256 \times 200$ equidistant points. For training, $N_{ic} = 100$ initial condition points and $N_{bc} = 100$ boundary condition points are randomly generated, while $N_{r} = 10,000$ residual points are sampled using the LHS technique. An enhanced PINN model with $5$ hidden layers and $40$ neurons per layer is then trained using these $10,000$ residual points. The $\mathbb{L}^2$ relative errors for the solutions $\psi_{1}$ and $\psi_{2}$ are reduced from $6.4\%$ and $6.4\%$ to $0.29\%$ and $0.28\%$, respectively, by using the Adam optimizer and then continuing the optimization process with the L-BFGS optimizer. In Fig.~\ref{fig:6}, the density plots of the exact and predicted soliton dynamics are shown in the top and middle rows, respectively. The top row demonstrates that the model successfully captures the m-shaped dynamics of the anti-dark one-soliton solution, which exhibits an m-shaped pulse with two smooth crests in both the $x-t$ and $y-t$ planes. Notably, the crests in the $x-t$ plane are wider than those in the $y-t$ plane. In the $x-y$ plane, shown in the middle row, the pulse propagates along a straight line. The bottom row presents a comparison between the exact and predicted dynamics at different temporal levels for $y = 4$ and $x = 25$, highlighting the accuracy of the trained model. The evolution of the m-shaped anti-dark one-soliton solution and the corresponding point-wise errors in the $x-t$ and $y-t$ planes are depicted in Fig.~\ref{fig:7} for $x = 0$ and $y = 0$ through three-dimensional plots. In Fig.~\ref{fig:8}, the loss curves during training with $30,000$ iterations of the Adam optimizer and $5,170$ iterations of the L-BFGS optimizer are illustrated. As expected, the Adam optimizer exhibits a slow decrease in loss values with many oscillations during training. From the middle panel of Fig.~\ref{fig:8}, the MSE loss for each individual residual equation oscillates relatively close to one another. On the other hand, the loss values exhibit smooth behavior without any oscillations when using the L-BFGS optimizer, as shown in the right panel of Fig.~\ref{fig:8}. This indicates that the L-BFGS optimizer provides a more stable and effective framework for training the PINN model in this context. Table \ref{tab3} shows that the proposed enhanced PINN method achieves the lowest $\mathbb{L}^{2}$ relative errors of $0.29\%$ for $\psi_{1}$ and $0.28\%$ for $\psi_{2}$, outperforming both the method in \cite{bib77} and the standard PINN. All networks have the same structure and we set $N_{r} = 16,200$ for the method in \cite{bib77} and the standard PINN. The enhanced PINN attains the lowest loss values by optimizing the network parameters with the Adam optimizer, followed by the L-BFGS optimizer. Furthermore, its training time is shorter than that of the method in \cite{bib77} and similar to the standard PINN despite its much higher accuracy. This efficiency is due to its region-specific training approach with fewer residual points and different distributions per equation residual, whereas the other methods use larger residual point datasets with the same distribution, increasing computational cost. Fig.~\ref{fig:88} illustrates the evolution of the $\mathbb{L}^{2}$ relative error of $\psi_{1}$ for training the enhanced PINN model with varying network depth (left panel) and width (right panel). As shown in the left panel, increasing the network depth initially reduces the $\mathbb{L}^{2}$ relative error, but further increases lead to slight error growth, suggesting overfitting or optimization challenges with very deep architectures. In contrast, increasing the network width (right panel) consistently decreases the error up to a certain width, after which it stabilizes. Fig.~\ref{fig:888} shows the effect of the number of residual points $N_{r}$, initial condition points $N_{ic}$, and boundary condition points $N_{bc}$ on the $\mathbb{L}^{2}$ relative error of $\psi_{1}$ for training the enhanced PINN model. Increasing $N_{r}$ (left panel) significantly reduces the error, indicating its strong influence on enforcing the PDE. Increasing $N_{ic}$ (middle panel) also decreases the error, improving initial condition enforcement. In contrast, increasing $N_{bc}$ (right panel) has a smaller impact on error reduction. Overall, the number of residual points has the greatest effect on model accuracy.
\begin{table}[h!]
    \centering
    \caption{Comparison results in terms of the $\mathbb{L}^{2}$ relative error, loss value, and training time (s).}
    \begin{tabular}{cccccc}
    \toprule
    &$\mathbb{L}^{2}$ error&&Loss value&&\\
    \cmidrule(r){2-3} \cmidrule(r){4-5} 
    Method&$\psi_{1}$&$\psi_{2}$ &Adam&L-BFGS&Training time (s)\\
    \midrule
    Enhanced PINN&$0.29\%$&$0.28\%$&$1.5215e-02$&$8.8810e-03$&$2152.5738$\\
    Method in \cite{bib77}&$8.03\%$&$8.05\%$&$2.0385e-00$&$8.8399e-01$&$2403.9105$\\
    PINN&$15.19\%$&$15.18\%$&$9.8001e+01$&$7.0091e-00$&$2152.5801$\\
    \bottomrule
    \end{tabular}
    \label{tab3}
\end{table}
\subsubsection{Discovery of two-soliton solutions}

The general form of the dark two-soliton solutions for the system of equations \eqref{eq:3} can be obtained by applying the Hirota method as follows \cite{bib57}:
\begin{align}
\psi_{1} &= \mu e^{ib(t)} \frac{1 + e^{\xi_{1}(x,y,t) + i2\theta_{1}}+e^{\xi_{2}(x,y,t) + i2\theta_{2}} + \hat{B}e^{\xi_{1}(x,y,t) + \xi_{2}(x,y,t) + i2\theta_{1}+i2\theta_{2}}}{1 + e^{\xi_{1}(x,y,t)} + e^{\xi_{2}(x,y,t)} + \hat{A}e^{\xi_{1}(x,y,t) + \xi_{2}(x,y,t)}}, \label{eq:17}\\
\psi_{2} &= \lambda e^{ib(t)} \frac{1 + e^{\xi_{1}(x,y,t) + i2\theta_{1}}+e^{\xi_{2}(x,y,t) + i2\theta_{2}} + \hat{C}e^{\xi_{1}(x,y,t) + \xi_{2}(x,y,t) + i2\theta_{1}+i2\theta_{2}}}{1 + e^{\xi_{1}(x,y,t)} + e^{\xi_{2}(x,y,t)} + \hat{A}e^{\xi_{1}(x,y,t) + \xi_{2}(x,y,t)}}, \label{eq:18}
\end{align}
where
\begin{align}
k &:= \eta(\lvert \mu \rvert^2 + \lvert \lambda \rvert^2), \quad \sigma_{2} := \frac{\lvert \sin(\theta_{2}) \rvert}{\lvert \sin(\theta_{1}) \rvert }\sigma_{1}, \quad b(t) := -k\int \alpha(t) dt,\\
\rho_{1} &:= \sqrt{2k\sin^2(\theta_{1}) - \sigma_{1}^2}, \quad \rho_{2} := \sqrt{2k\sin^2(\theta_{2}) - \sigma_{2}^2},\\
w_{1}(t) &:= k\sin(2\theta_{1})\int \alpha(t)dt, \quad w_{2}(t) := k\sin(2\theta_{2})\int \alpha(t)dt, \\
\hat{A} = \hat{B} = \hat{C} &:= \frac{\sigma_{1}\sigma_{2} + \rho_{1}\rho_{2} - 2k\sin(\theta_{1})\sin(\theta_{2})\cos(\theta_{1} - \theta_{2})}{\sigma_{1}\sigma_{2} + \rho_{1}\rho_{2} - 2k\sin(\theta_{1})\sin(\theta_{2})\cos(\theta_{1} + \theta_{2})},\\
\xi_{1}(x,y,t) &:= \sigma_{1}x + \rho_{1}y + w_{1}(t), \quad \xi_{2}(x,y,t) := \sigma_{2}x + \rho_{2}y + w_{2}(t). \label{eq:23}
\end{align}

Here, $\eta$, $\sigma_{1}$, $\sigma_{2}$, $\rho_{1}$, $\rho_{2}$, $\hat{A}$, $\hat{B}$, and $\hat{C}$ are all real constants, $\mu$ and $\lambda$ are complex constants, and $\theta_{1}$ and $\theta_{2}$ are the real ones and can be varied within the range $[0, 2\pi]$. $w_{1}(t)$ and $w_{2}(t)$ are real functions of the temporal variable $t$. It has been investigated that the amplitudes of the two circularly polarized waves are affected by $\mu$ and $\lambda$. 

\noindent\textbf{Case 3:} Parallel dark two-solitons \label{C:3}

\begin{figure}[h!]
\centering
\includegraphics[height = 9cm,width = 13cm]{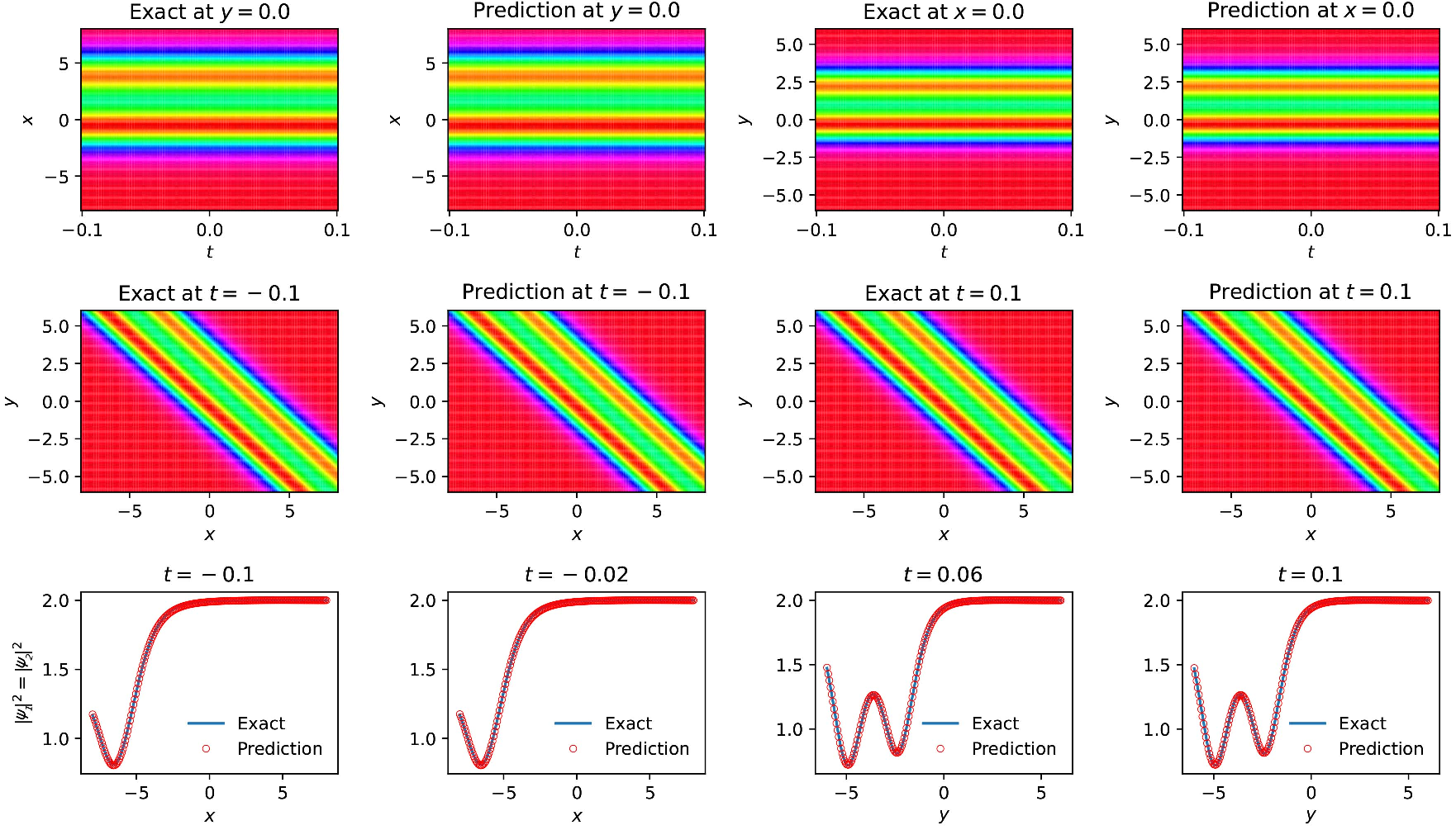}
\caption{The density plots of the exact and predicted dynamics (top and middle rows) and the comparison between the exact and predicted solutions at different temporal levels (bottom row).}
\label{fig:9}
\end{figure}

\begin{figure}[h!]
\centering
\includegraphics[height = 3cm,width = 13cm]{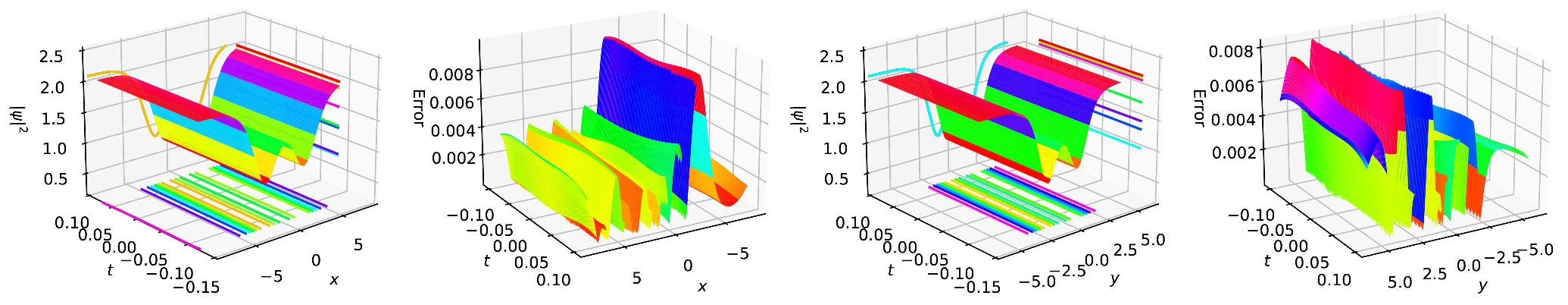}
\caption{Three-dimensional plots with contour maps and the corresponding point-wise errors for $\psi_{j}$ in the $x-t$ plane (first and second panels) and $y-t$ plane (third and fourth panels), from left to right.}
\label{fig:10}
\end{figure}

\begin{figure}[h!]
\centering
\includegraphics[height = 3cm,width = 13cm]{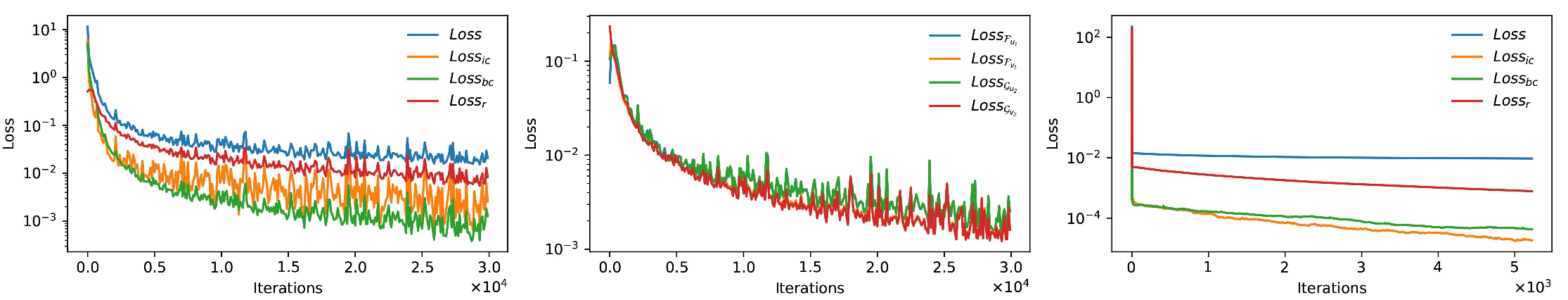}
\caption{The loss curves during the training with $30,000$ iterations of the Adam optimizer (left and middle panels) and the loss curves with $5,215$ iterations of the L-BFGS optimizer (right panel).}
\label{fig:11}
\end{figure}

By setting the perturbation coefficient $\alpha(t) = \frac{1}{2}t$, and choosing the other constants as $\eta = 1, \sigma_{1} = 1, k = 4, \mu = \lambda = 1 + i, \theta_{1} = \frac{\pi}{4}$, and $\theta_{2} = \frac{\pi}{3}$ in \eqref{eq:17}-\eqref{eq:23}, one can obtain the parallel dark two-soliton solutions to the (2+ 1)-dimensional VC-CNLSE \eqref{eq:3}. The nonlinearity coefficient can be considered proportional to the dispersion coefficient, expressed as $\beta(t) = -\eta \alpha(t)$. The spatial and temporal domains are set as $\Omega = [-8, 8] \times [-6, 6]$ and $[-0.1, 0.1]$, discretized into $256 \times 256 \times 201$ equidistant points. We set $N_{ic} = 200$ and $N_{bc} = 200$, and generate $N_{r} = 10,000$ residual points randomly using the LHS strategy. We aim to recover the dark two-soliton solutions given by \eqref{eq:17} and \eqref{eq:18} by applying the enhanced PINN framework. In doing so, we use a neural network model with $4$ hidden layers and $100$ neurons per layer. After $30,000$ iterations of the Adam optimizer with learning rate $0.0001$, followed by $5,215$ iterations of the L-BFGS optimizer the $\mathbb{L}^2$ relative errors for $\psi_{1}$ and $\psi_{2}$ are reduced from $0.31\%$ and $0.32\%$ to $0.22\%$ and $0.25\%$, respectively. In Fig.~\ref{fig:9}, the density plots of the exact and predicted dynamics in the $x-t$ and $y-t$ planes (top row) and the $x-y$ plane (middle row) are displayed, confirming that the parallel dark two-solitons propagate without interaction. Additionally, we compare the exact and predicted solutions at different temporal levels in the bottom row of Fig.~\ref{fig:9}, which shows that our trained model successfully captured the data-driven solutions to the system of equations \eqref{eq:3}. The evolution of the predicted dynamics in the $x-t$ plane (for $y = 6$) and the $y-t$ plane (for $x = 8$), along with the corresponding point-wise errors, are displayed in Fig.~\ref{fig:10}. The behavior of the loss functions during training with the Adam optimizer and the L-BFGS optimizer is illustrated in Fig.~\ref{fig:11}. In the left panel, the loss trajectories through the optimization process using the Adam optimizer are illustrated, showing a slow decrease in loss values with many oscillations. From the middle panel, one can see that the individual residual losses behave relatively close to each other. In the right panel of Fig.~\ref{fig:11}, the L-BFGS optimizer exhibits smooth behavior of the loss values during training. In Table~\ref{tab4}, a comparison between the proposed enhanced PINN and the method in \cite{bib77} and the standard PINN is presented. All networks have the same structure, and $N_{r} = 16,200$ is set for the two other methods. Additionally, the method in \cite{bib77} uses the same activation functions as the enhanced PINN. One can see that, the enhanced PINN achieves the lowest $\mathbb{L}^{2}$ relative errors ($0.22\%$ for $\psi_{1}$ and $0.25\%$ for $\psi_{2}$) and the lowest loss values, with a training time ($3010$ s) shorter than the method in \cite{bib77} $(3393$ s) and comparable to the standard PINN. In contrast, both other methods show significantly higher errors and losses.
\begin{table}[h!]
    \centering
    \caption{Comparison results in terms of the $\mathbb{L}^{2}$ relative error, loss value, and training time (s).}
    \begin{tabular}{cccccc}
    \toprule
    &$\mathbb{L}^{2}$ error&&Loss value&&\\
    \cmidrule(r){2-3} \cmidrule(r){4-5} 
    Method&$\psi_{1}$&$\psi_{2}$ &Adam&L-BFGS&Training time (s)\\
    \midrule
    Enhanced PINN&$0.22\%$&$0.25\%$&$2.7511e-02$&$8.1037e-03$&$3010.3263$\\
    Method in \cite{bib77}&$10.51\%$&$12.15\%$&$9.6591e-00$&$1.7260e-00$&$3392.7019$\\    PINN&$17.62\%$&$19.03\%$&$15.0093e+01$&$9.1101e-00$&$3011.9031$\\
    \bottomrule
    \end{tabular}
    \label{tab4}
\end{table}

\noindent\textbf{Case 4:} Parallel cubic dark two-solitons \label{C:4}

\begin{figure}[h!]
\centering
\includegraphics[height = 9cm,width = 13cm]{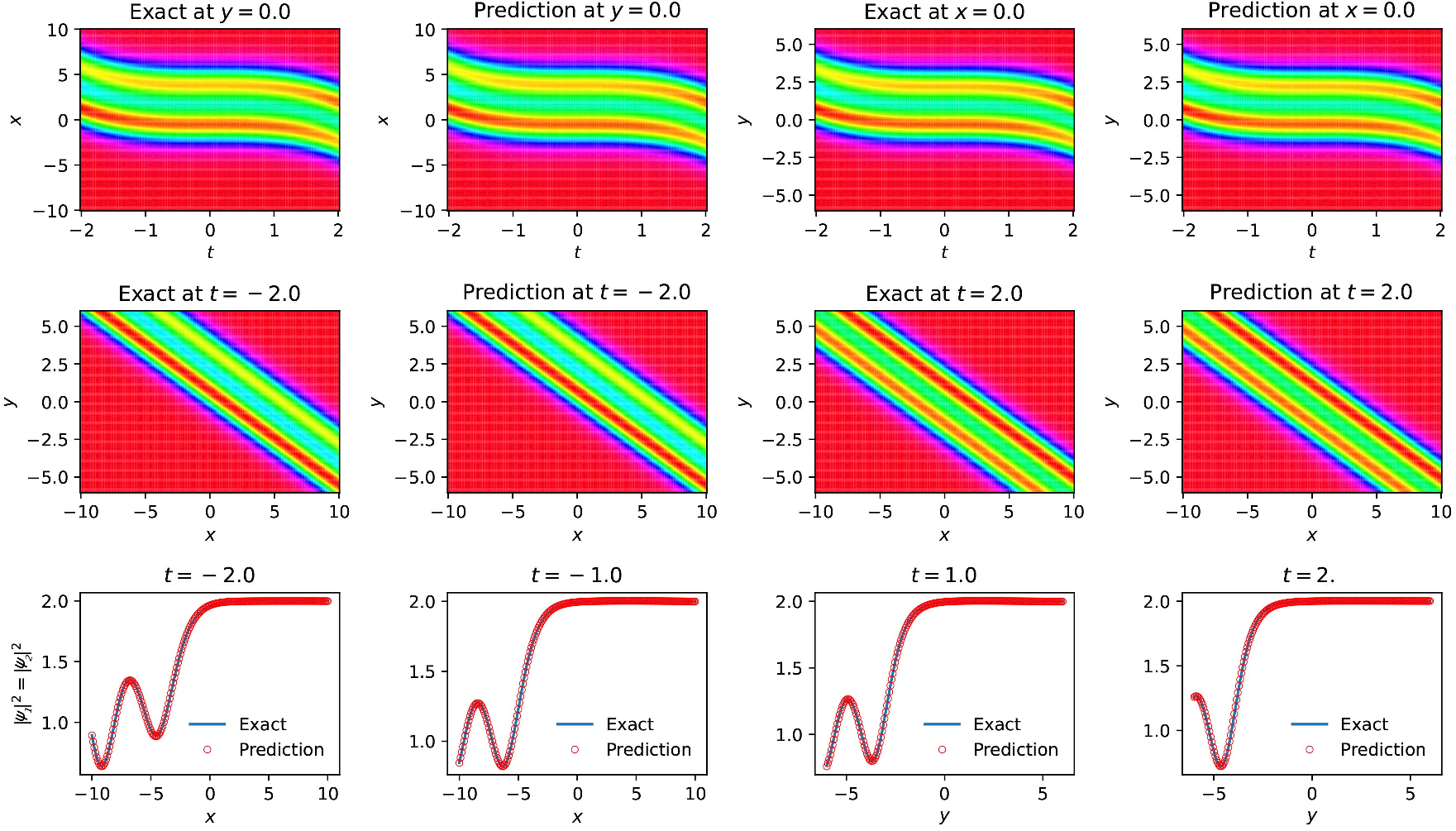}
\caption{The density plots of the exact and predicted dynamics (top and middle rows) and the comparison between the exact and predicted solutions at different temporal levels (bottom row).}
\label{fig:12}
\end{figure}

\begin{figure}[h!]
\centering
\includegraphics[height = 3cm,width = 13cm]{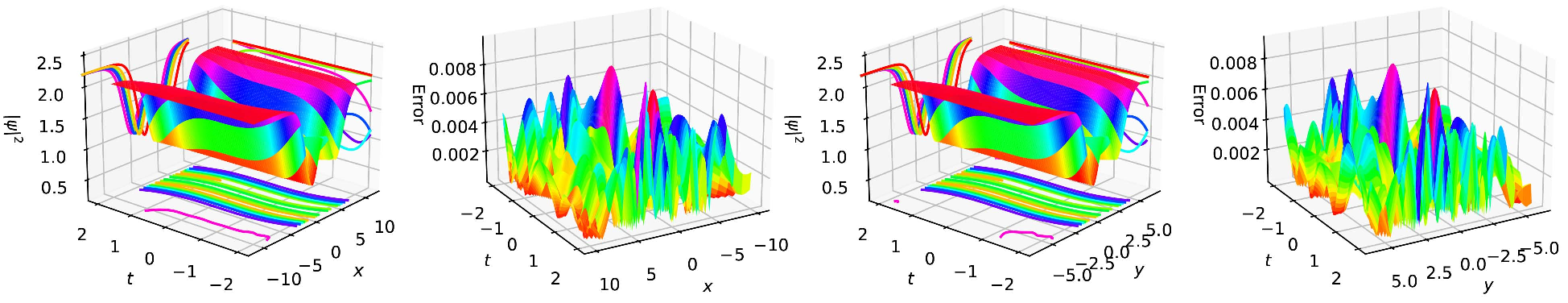}
\caption{Three-dimensional plots with contour maps and the corresponding point-wise errors for $\psi_{j}$ in the $x-t$ plane (first and second panels) and $y-t$ plane (third and fourth panels), from left to right.}
\label{fig:13}
\end{figure}

\begin{figure}[h!]
\centering
\includegraphics[height = 3cm,width = 13cm]{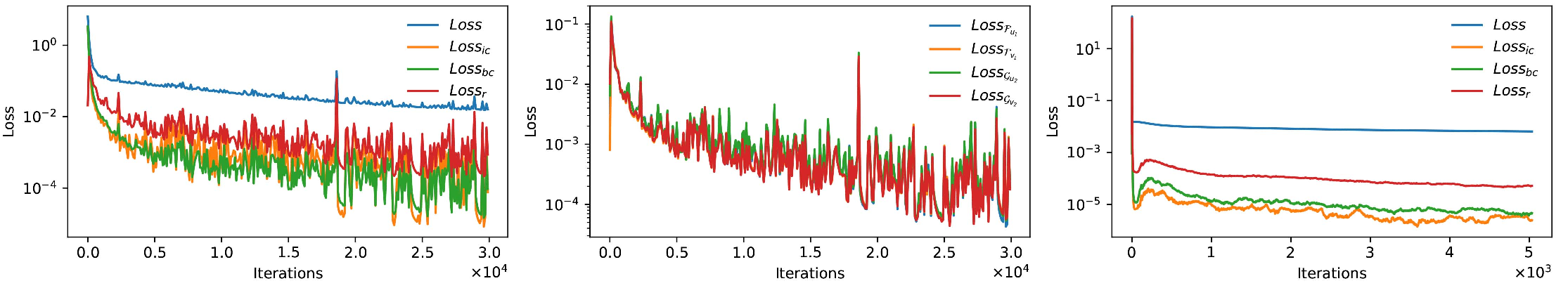}
\caption{The loss curves during the training with $30,000$ iterations of the Adam optimizer (left and middle panels) and the loss curves with $5,034$ iterations of the L-BFGS optimizer (right panel).}
\label{fig:14}
\end{figure}

By taking $\eta = 1, \sigma_{1} = 1, k = 4, \mu = \lambda = 1 + i, \theta_{1} = \frac{\pi}{4}, \theta_{2} = \frac{\pi}{3}$, and choosing the dispersion coefficient as $\alpha(t) = \frac{1}{5}t^2$, the cubic dark two-dolitons to the (2+ 1)-dimensional VC-CNLSE \eqref{eq:3} can be obtained. Here, the nonlinearity coefficient is considered as $\beta(t) = -\eta\alpha(t)$. 
As before, we first divide the computational domain $ [-10, 10]\times [-6,6]\times[-2, 2]$ into $256 \times 256 \time 201$ discrete equidistant points. We set $N_{ic} = 200$ and $N_{bc} = 200$ to be randomly chosen. We use the LHS strategy to generate $N_{r} = 10,000$ residual points within the computational domain. Our enhanced PINN model comprises a network with $4$ hidden layers and $100$ neurons in each layer. After $30,000$ iterations of the Adam optimizer, followed by $5,034$ iterations of the L-BFGS optimizer the $\mathbb{L}^{2}$ relative errors for $\psi_{1}$ and $\psi_{2}$ are measured from $0.64\%$ and $0.65\%$ to $0.15\%$ and $0.16\%$, respectively. The dynamical behaviors of the exact and predicted cubic dark two-solitons are illustrated in the $x-t$ and $y-t$ planes in the top row, and the $x-y$ plane in the middle row of Fig.~\ref{fig:12}. One can see that the dark two-solitons do not interact with each other during propagation. Comparisons between predicted dynamics at different temporal levels are presented in the bottom row of Fig.~\ref{fig:12}. In Fig.~\ref{fig:13}, the evolution of the predicted dynamics in the $x-t$ plane (for $y = 6$) and the $y-t$ plane (for $x = 10$), with the corresponding point-wise errors, is displayed through three-dimensional plots. The loss values during the training phase with the Adam optimizer and L-BFGS optimizer are illustrated in Fig.~\ref{fig:14}. It can be seen that, the Adam optimizer exhibits oscillatory behavior in the loss functions during training, whereas the L-BFGS optimizer shows gradual descent without such oscillations. In the middle panel of Fig.~\ref{fig:14}, the behavior of the individual residual losses is depicted, showing that these loss values are relatively close to one another. In Table~\ref{tab5}, the performance of the proposed enhanced PINN is compared with the method in \cite{bib77} and the standard PINN. All networks have the same structure, and $N_{r} = 16,200$ is set for the other two methods. Additionally, the method in \cite{bib77} uses the same activation functions as the enhanced PINN. The enhanced PINN achieves the lowest $\mathbb{L}^{2}$ relative errors and the lowest loss values for both Adam and L-BFGS optimizers. Its training time is also shorter than that of the method in \cite{bib77} and similar to the standard PINN. In contrast, both the method in \cite{bib77} and the standard PINN result in significantly higher errors and loss values.
\begin{table}[h!]
    \centering
    \caption{Comparison results in terms of the $\mathbb{L}^{2}$ relative error, loss value, and training time (s).}
    \begin{tabular}{cccccc}
    \toprule
    &$\mathbb{L}^{2}$ error&&Loss value&&\\
    \cmidrule(r){2-3} \cmidrule(r){4-5} 
    Method&$\psi_{1}$&$\psi_{2}$ &Adam&L-BFGS&Training time (s)\\
    \midrule
    Enhanced PINN&$0.15\%$&$0.16\%$&$1.6082e-02$&$6.5041e-03$&$3005.8592$\\
    Method in \cite{bib77}&$9.62\%$&$9.61\%$&$4.27301e-00$&$5.0026e-01$&$3209.9162$\\    PINN&$13.04\%$&$13.05\%$&$5.28301e+01$&$8.2830e-01$&$3005.7590$\\
    \bottomrule
    \end{tabular}
    \label{tab5}
\end{table}
\subsection{Data-driven parameter discovery to (2+ 1)-dimensional VC-CNLSE}

In this part of the paper, we consider the data-driven parameter discovery for the (2+ 1)-dimensional VC-CNLSE \eqref{eq:3}. When solving the inverse problems, we do not need to make significant changes to the code used for solving the forward problems. We investigate two approaches for parameter discovery in the system of model equations \eqref{eq:3}: constant parameter discovery and variable parameter discovery. 

$\bullet$ Data-driven constant parameter discovery

\begin{figure}[h!]
\centering
\includegraphics[height = 3cm,width = 13cm]{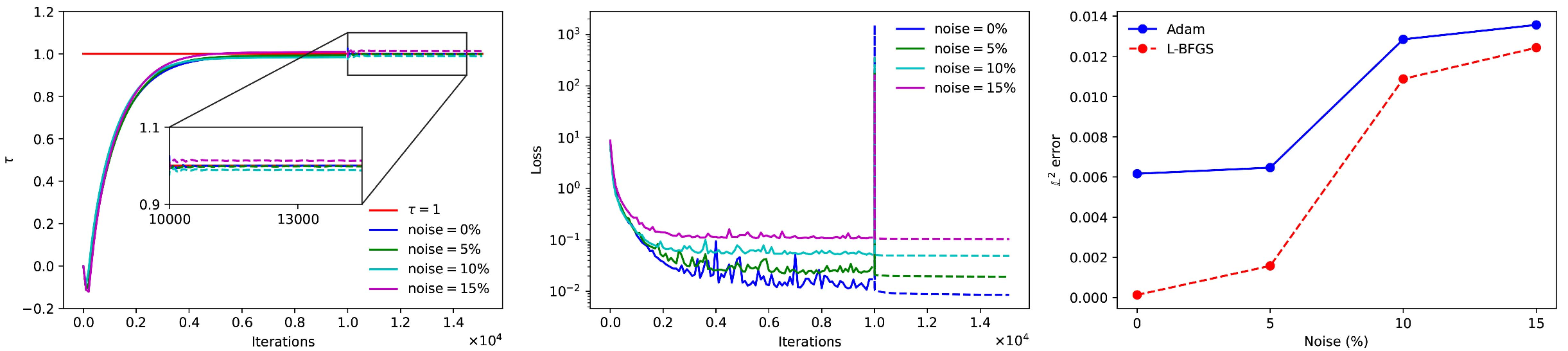}
\caption{The predictions of the constant coefficient $\tau$ (left panel), the behavior of the loss functions Adam + L-BFGS (middle panel), and the prediction errors (right panel) under different noise intensities.}
\label{fig:15}
\end{figure}

For the constant parameter discovery, we use the introduced enhanced PINN framework to predict the constant coefficient $\tau$ through training a PINN model. Given the same parameters for the parabolic vector dark one-soliton, we discretize the spatial and temporal domains into $256 \times 256 \times 201$ equidistant points. We randomly sample $N_{o} = 1,000$ points from the generated data, then create a set of observations of the exact solutions $\Big{\{}u_{1}^{k}, v_{1}^{k}, u_{2}^{k}, v_{2}^{k}\Big{\}}_{k = 1}^{N_{o}}$ based on the sampled data points. We sample $N_{r} = 10,000$ initial residual points. The data-driven parameter discovery to predict the constant coefficient $\tau$ will be performed by training an enhanced PINN model, which is equipped with a neural network containing $5$ hidden layers with $40$ neurons per layer, through an optimization process that minimizes the following loss function:
\begin{align*}
Loss(\boldsymbol{\theta}) &:= \frac{1}{N_{o}} \sum_{k = 1}^{N_{o}} \Big{(} \Big{\lvert} u_{1}^{k} - \hat{u}_{1, \boldsymbol{\theta}}(\bold{x}^{k}, t^{k}) \Big{\rvert}^2 +  \Big{\lvert} v_{1}^{k} - \hat{v}_{1, \boldsymbol{\theta}}(\bold{x}^{k}, t ^{k}) \Big{\rvert}^2 \\
&+ \Big{\lvert} u_{2}^{k} - \hat{u}_{2, \boldsymbol{\theta}}(\bold{x}^{k}, t^{k}) \Big{\rvert}^2 +  \Big{\lvert} v_{2}^{k} - \hat{v}_{2, \boldsymbol{\theta}}(\bold{x}^{k}, t^{k}) \Big{\rvert}^{2}\Big{)}\\
&+\frac{1}{N_{r}} \sum_{k = 1}^{N_{r}}\Big{(} w_{r}^{\mathcal{F}_{\hat{u}}}\Big{\lvert} \mathcal{F}_{\hat{u}_{1,\boldsymbol{\theta}}}(\bold{x}^{k}, t^{k}) \Big{\rvert}^{2} + w_{r}^{\mathcal{F}_{\hat{v}}}\Big{\lvert} \mathcal{F}_{\hat{v}_{1,\boldsymbol{\theta}}}(\bold{x}^{k}, t^{k}) \Big{\rvert}^{2} \\
&+ w_{r}^{\mathcal{G}_{\hat{u}}}\Big{\lvert} \mathcal{G}_{\hat{u}_{2,\boldsymbol{\theta}}}(\bold{x}^{k}, t^{k}) \Big{\rvert}^{2} + w_{r}^{\mathcal{G}_{\hat{v}}}\Big{\lvert} \mathcal{G}_{\hat{v}_{2,\boldsymbol{\theta}}}(\bold{x}^{k}, t^{k}) \Big{\rvert}^{2} \Big{)} + w_{a}\cdot Loss_{a}(\boldsymbol{\theta}),
\end{align*}
where $\boldsymbol{\theta}$ represents the set of all trainable parameters, including the weights, biases, scalable parameters, and the parameter $\hat{\tau}$, which serves as a prediction of the true value of $\tau$. We initialized this trainable parameter as $\hat{\tau} = 0$. Minimizing this loss function allows our PINN model to align its predictions with data-driven observations while embedding the physical laws into the model's knowledge. We allow our model to learn the true value of the constant coefficient $\tau$ by treating this parameter as one of the trainable parameters within the underlying neural network model. We used $10,000$ iterations of the Adam optimizer with the default learning rate, followed by a run of the L-BFGS optimizer, to obtain the optimal values of the trainable parameters. During this process, the $\mathbb{L}^{2}$ relative error between the true value of $\tau$ and its prediction $\hat{\tau}$ decreased from $0.61\%$ to $0.013\%$, demonstrating that our trained model can predict the true value of $\tau$ with high accuracy. To evaluate the performance of the enhanced PINN model for data-driven parameter discovery in the (2+ 1)-dimensional VC-CNLSE under noisy conditions, we introduce different noise intensities to the observational data. For a fair comparison, we use the same initialization for all networks trained with different noise intensities. As reported in Table~\ref{tab:1}, the enhanced PINN model accurately predicts the true value of the constant coefficient $\tau$ even when the observational data is corrupted with noise. It is evident that increasing noise levels leads to higher prediction errors and negatively impacts loss convergence. The left panel of Fig.~\ref{fig:15} depicts the process of our PINN model learning the true value of the constant coefficient $\tau$. During the iterations of the Adam optimizer (solid lines), followed by the L-BFGS optimizer (dashed lines), our PINN model gradually predicts the true value of $\tau$ with high accuracy. One can see that increasing the noise levels will have a negative impact on the predictions. Training under different noise intensities alters the loss values, as shown in the middle panel of Fig.~\ref{fig:15}. The $\mathbb{L}^2$ relative errors between the constant coefficient $\tau$ and its prediction $\hat{\tau}$ for different levels of added noise are illustrated in the right panel of Fig.~\ref{fig:15}. It can be seen that as the noise level increases from $0$ to $15$, the relative error also increases, and continuing the optimization with the L-BFGS optimizer does not have a significant effect on prediction accuracy. In Tabel~\ref{tab:2}, we investigate the effect of the number of observations on the prediction accuracy through the lens of $\mathbb{L}^{2}$ relative error. It can be seen that increasing the number of observations reduces the relative error, even in the presence of noisy data. In summary, our enhanced PINN model achieves lower prediction error and faster convergence with clean data (noise $= 0\%$). However, introducing noise into the data negatively impacts the model's prediction accuracy and convergence.
\begin{table}[h!]
\caption{The $\mathbb{L}^2$ relative errors and loss values of the discovered coefficient under noise intensities (Adam+L-BFGS).}
\begin{tabular}{cccc}
\toprule
Noise ($\%$) &Predicted values of $\tau$ & $\mathbb{L}^2$ relative errors ($\%$)&Loss values\\
\midrule
$0$&$0.9998723$&$0.013$&$8.536e-03$\\
$5$&$0.9984364$&$0.16$&$1.934e-02$\\
$10$&$0.98912466$&$1.09$&$4.896e-02$\\
$15$&$1.01243$&$1.24$&$1.042e-01$\\
\bottomrule
\end{tabular}
\label{tab:1}
\end{table}
\begin{table}[h!]
    \centering
    \caption{The $\mathbb{L}^{2}$ relative errors of $\tau$ under noise intensities and varying $N_{o}$ (Adam + L-BFGS).}
    \begin{tabular}{lccccc}
    \toprule
    &$\mathbb{L}^{2}$ relative errors ($\%$) & & &&  \\
    \cmidrule(r){2-5}
    Data type&$N_{o} = 100$& $500$ & $1000$ & $1500$& \\
    \midrule
   Clean data ($0\%$) &$0.045$& $0.0027$ & $0.013$ & $0.0013$& \\
   Noisy data ($5\%$) &$1.57$& $0.45$ & $0.056$ & $0.11$& \\
   Noisy data ($10\%$) &$0.27$& $1.01$ & $1.09$ & $0.38$& \\
   Noisy data ($15\%$) &$11.93$& $1.20$ & $1.24$ & $0.17$& \\
    \bottomrule
    \end{tabular}
    \label{tab:2}
\end{table}

$\bullet$ Data-driven variable parameter discovery

Here, we investigate the capability of the proposed enhanced PINN framework in capturing the dynamic behavior of the variable dispersion and nonlinearity coefficients $\alpha(t)$ and $\beta(t)$ for the (2+ 1)-dimensional VC-CNLSE \eqref{eq:3}. In doing so, we employ a dual-network strategy for the enhanced PINN algorithm, where one network is responsible for predicting the solutions $\psi_{1}$ and $\psi_{2}$, while another network is tasked with predicting the dynamic behavior of the variable coefficients. Incorporating the physical laws described by the system of equations \eqref{eq:3} into the training phase is the key feature of the enhanced PINN framework, enabling the model to learn the dynamic behavior of the variable coefficients with a small amount of training data. We consider the same parameter configuration as for the vector m-shaped anti-dark one-soliton solution and train a PINN model based on the enhanced framework to predict the dynamic behavior of the following dispersion and nonlinearity coefficients:
\begin{equation*}
\alpha(t) = t \ln(t^2), \quad \beta(t) = -\frac{\csc^{2}(\theta)(k^2 + m^2)t \ln(t^2)}{2(\lvert \lambda \rvert^2 + \lvert \mu \rvert^2)}.
\end{equation*}

\begin{figure}[h!]
\centering
\includegraphics[height = 3cm,width = 13cm]{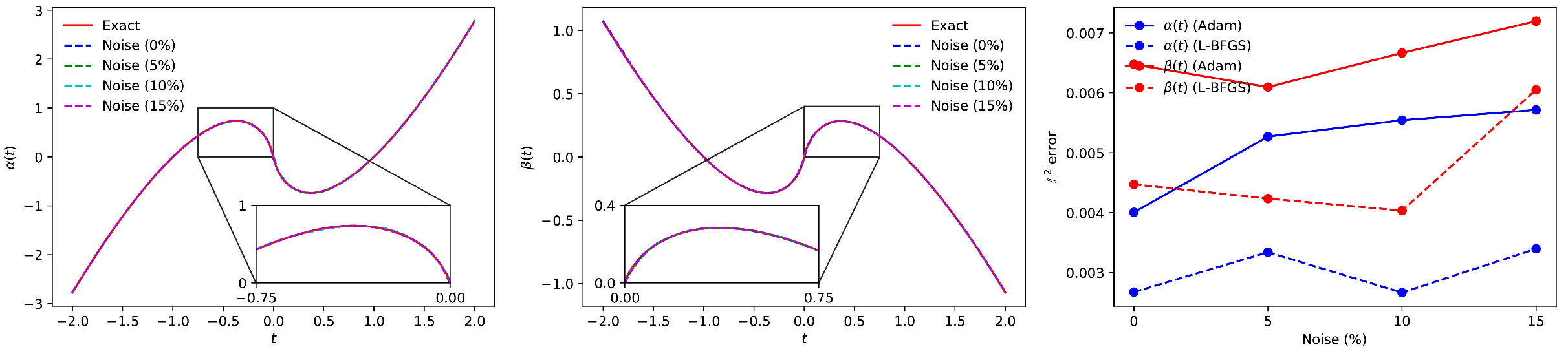}
\caption{The predictions of the dispersion coefficient $\alpha(t)$ (left panel) and the nonlinearity coefficient $\beta(t)$ (middle panel), and the related prediction errors (right panel) under different noise intensities.}
\label{fig:16}
\end{figure}

Since these coefficients are functions of the temporal variable $t$, the corresponding neural network should process the temporal coordinate as input data to learn the dynamics of these coefficients. We divide the computational domain $\Omega \times [-2, 2]$ into $256 \times 256 \times 200$ discrete equidistant points and randomly sample $N_o = 2,000$ data observations from the generated data points. We also sample $N_r = 20,000$ residual points using the LHS strategy within the computational domain. Additionally, we divide the temporal domain $[-2, 2]$ into $1,000$ equidistant points to assess the accuracy of the underlying PINN model in predicting the variable coefficients through the lens of the $\mathbb{L}^{2}$ relative error. Our PINN model employs two neural networks: the first, parameterized by $\boldsymbol{\theta}_{1}$, takes spatial and temporal coordinates as inputs to train and predict the solutions $\psi_1$ and $\psi_2$ for forward problem solving; the second, parameterized by $\boldsymbol{\theta}_{2}$, uses only the temporal coordinate as input to predict the variable coefficients. Once these coefficients are predicted, they are directly integrated into the physics-informed components described in \eqref{eq:10}. Both networks may or may not have the same structure and will be trained simultaneously. In this experiment, the first network has $5$ hidden layers with $40$ neurons in each, while the second network has $3$ hidden layers with $30$ neurons per layer. Then, the underlying PINN model aims to learn the dynamic behavior of both the solutions $\psi_1$ and $\psi_2$, as well as the dynamic behavior of the variable coefficients, through a minimization problem by minimizing the following loss function:
\begin{align*}
Loss(\boldsymbol{\theta}_{1}, \boldsymbol{\theta}_{2}) &:= \frac{1}{N_{o}} \sum_{k = 1}^{N_{o}} \Big{(} \Big{\lvert} u_{1}^{k} - \hat{u}_{1, \boldsymbol{\theta}_{1}}(\bold{x}^{k}, t^{k}) \Big{\rvert}^2 +  \Big{\lvert} v_{1}^{k} - \hat{v}_{1, \boldsymbol{\theta}_{1}}(\bold{x}^{k}, t ^{k}) \Big{\rvert}^2 \\
&+ \Big{\lvert} u_{2}^{k} - \hat{u}_{2, \boldsymbol{\theta}_{1}}(\bold{x}^{k}, t^{k}) \Big{\rvert}^2 +  \Big{\lvert} v_{2}^{k} - \hat{v}_{2, \boldsymbol{\theta}_{1}}(\bold{x}^{k}, t^{k}) \Big{\rvert}^{2}\Big{)}\\
&+\frac{1}{N_{r}} \sum_{k = 1}^{N_{r}}\Big{(} w_{r}^{\mathcal{F}_{\hat{u}}}\Big{\lvert} \mathcal{F}_{\hat{u}_{1,\boldsymbol{\theta}_{1},\boldsymbol{\theta}_{2}}}(\bold{x}^{k}, t^{k}) \Big{\rvert}^{2} + w_{r}^{\mathcal{F}_{\hat{v}}}\Big{\lvert} \mathcal{F}_{\hat{v}_{1,\boldsymbol{\theta}_{1},\boldsymbol{\theta}_{2}}}(\bold{x}^{k}, t^{k}) \Big{\rvert}^{2} \\
&+ w_{r}^{\mathcal{G}_{\hat{u}}}\Big{\lvert} \mathcal{G}_{\hat{u}_{2,\boldsymbol{\theta}_{1},\boldsymbol{\theta}_{2}}}(\bold{x}^{k}, t^{k}) \Big{\rvert}^{2} + w_{r}^{\mathcal{G}_{\hat{v}}}\Big{\lvert} \mathcal{G}_{\hat{v}_{2,\boldsymbol{\theta}_{1},\boldsymbol{\theta}_{2}}}(\bold{x}^{k}, t^{k}) \Big{\rvert}^{2} \Big{)}\\&+\frac{1}{N_{vc}}\sum_{k = 1}^{N_{vc}} {\Big (} {\Big \lvert} \alpha_{\boldsymbol{\theta}_{2}}(t_{k}) - \alpha(t_{k}){\Big \rvert}^{2} + {\Big \lvert} \beta_{\boldsymbol{\theta}_{2}}(t_{k}) - \beta(t_{k}){\Big \rvert}^{2} {\Big)}\\
&+ w_{a_{1}}\cdot Loss_{a_{1}}(\boldsymbol{\theta}_{1}) + w_{a_{2}}\cdot Loss_{a_{2}}(\boldsymbol{\theta}_{2}),
\end{align*}
where $\boldsymbol{\theta}_{1}$ and $\boldsymbol{\theta}_{2}$ are the sets of trainable parameters for the first network, which outputs the real and imaginary parts of the decomposed solutions $\psi_{1}$ and $\psi_{2}$, and the second network, which outputs the predictions of the variable coefficients, respectively. $Loss_{a_{1}}$ and $Loss_{a_{2}}$ denote the slope recovery loss terms for the first and second networks, respectively, and $w_{a_{1}}$ and $w_{a_{2}}$ are the corresponding weight coefficients. For this example, we set $w_{a_{1}} = 1e-02$ and $w_{a_{2}} = 1e-01$. The set $\Big{\{}t_{k},\alpha(t_{k}),\beta(t_{k})\Big{\}}_{k = 1}^{N_{vc}}$ denotes the set of training data related to the boundary of the temporal interval. After $10,000$ iterations of the Adam optimizer and some iterations of the L-BFGS optimizer with clean data (noise = $0\%$), the $\mathbb{L}^{2}$ relative error of the variable coefficients $\alpha(t)$ and $\beta(t)$ decreased from $0.40\%$ to $0.28\%$ and $0.65\%$ to $0.45\%$, respectively. When we add noise levels of $5\%$, $10\%$, and $15\%$ into the training observations, the $\mathbb{L}^{2}$ relative error of the dispersion coefficient $\alpha(t)$ changes from $0.53\%$, $0.55\%$, and $0.57\%$ to $0.33\%$, $0.27\%$, and $0.33\%$, respectively, when we first use iterations of the Adam optimizer and then continue the optimization process with the L-BFGS optimizer. This scenario holds for the nonlinearity coefficient $\beta(t)$, where the $\mathbb{L}^{2}$ relative error changes from $0.61\%$, $0.67\%$, and $0.72\%$ to $0.42\%$, $0.40\%$, and $0.60\%$, when we add noise levels of $5\%$, $10\%$, and $15\%$ into the training observations, respectively. The predictions of the dispersion and nonlinearity coefficients $\alpha(t)$ and $\beta(t)$ under different noise levels are displayed in the left and middle panels of Fig.~\ref{fig:16}. One can see that our enhanced trained PINN model can accurately predict the dynamic behavior of both variable coefficients, even in the presence of noise intensities. The $\mathbb{L}^{2}$ relative errors of both the dispersion and nonlinearity variable coefficients $\alpha(t)$ and $\beta(t)$ for different noise intensities are illustrated in the right panel of Fig.~\ref{fig:16}, which shows that increasing the noise levels does not significantly impact prediction accuracy. The obtained results show that our enhanced PINN model can accurately predict the behavior of the variable coefficients $\alpha(t)$ and $\beta(t)$ in the system of model equations \eqref{eq:3} with a small amount of training data.
\section{Conclusions}\label{sec:4}

In this study, we proposed an enhanced PINN framework for data-driven soliton solutions and parameter discovery to the (2+ 1)-dimensional VC-CNLSE. PINN algorithms enhance the capabilities of neural networks by incorporating the physical knowledge, described by differential equations, into the training phase. Incorporating physical knowledge into the training phase allows the network to learn from both data and the underlying principles governing the problem, leading to more accurate and physically consistent predictions. Our proposed framework uses a locally adaptive activation function to accelerate the convergence speed of the training process for the underlying PINN model. By leveraging the residual-based adaptive refinement (RAR) strategy, we introduced a region-specific weighted loss function, demonstrating that different distributions of physical knowledge are required when training a PINN model for the (2+ 1)-dimensional VC-CNLSE. The minimization problems are carried out using iterations of the Adam optimizer followed by iterations of the L-BFGS optimizer. The efficiency and accuracy of the proposed framework are evaluated in capturing the dynamics of vector dark and anti-dark one- and two-soliton structures, using various soliton solutions and data-driven predictions. The density plots of the exact and predicted solutions are displayed, and the related evolution dynamics are illustrated through three-dimensional plots. We also compared the exact and predicted solutions at different temporal levels to show the accuracy of the trained model's prediction. We performed data-driven parameter discovery for the (2+ 1)-dimensional VC-CNLSE, categorizing it into two classes: constant parameter discovery and variable parameter discovery. For constant parameter discovery, we employed an enhanced PINN model equipped with a single neural network to estimate the value of the constant parameter in the system of model equations. For variable parameter discovery, we used a dual-network strategy within the proposed PINN framework to capture the dynamic behavior of the variable dispersion and nonlinearity coefficients. As these coefficients are nonlinear functions of the temporal variable, this shows that our enhanced PINN framework is capable of capturing complex nonlinear variable coefficients. This robustness highlights the framework's potential for broader applications in complex optical fiber systems. The obtained results confirm that the proposed enhanced PINN framework offers a powerful and versatile approach for solving high-dimensional and complex solitonic dynamics in optical communication systems. Its robustness, accuracy, and adaptability make it a promising tool for advancing research and applications in nonlinear optics and related fields.

\section*{Declaration of competing interest}
The authors declare that there is no conflict of interest.

\section*{Data availability}
The data that support the findings of this study are available from the corresponding author upon reasonable request.

\end{document}